\documentclass[aps,tightenlines,twocolumn,floats,prd,nofootinbib,showpacs,preprintnumbers
]{revtex4-1}
\def\bea{\begin{eqnarray}}
\def\eea{\end{eqnarray}}
\def\bal{\begin{align}}
\def\eal{\end{align}}

\def\sfrac#1#2{{\textstyle \frac{#1}{#2}}}

\usepackage[dvips]{color}
\usepackage{graphics}
\usepackage{graphicx}
\usepackage{epsf}
\usepackage{amsmath}
\usepackage{amssymb}
\usepackage{bbold}
\usepackage{bm}
\usepackage{slashed}
\usepackage{float}
\usepackage{psfrag}
\usepackage{footnote}
\usepackage{xcolor}
\usepackage{textcomp}

\begin{document}

\preprint{CFTP/14-008}

\title{Electromagnetic rho-meson form factors in point-form relativistic quantum mechanics}


\author{Elmar P. Biernat}
\email[]{elmar.biernat@tecnico.ulisboa.pt}
\affiliation {Centro de F\'isica Te\'orica de Part\'iculas (CFTP),
Instituto Superior T\'ecnico (IST), Universidade de Lisboa,
Av. Rovisco Pais, 1049-001 Lisboa, Portugal}
\author{Wolfgang Schweiger}
\email[]{wolfgang.schweiger@uni-graz.at}
\affiliation{Institut f\"ur Physik, Fachbereich Theoretische Physik, Universit\"at Graz, A-8010 Graz, Austria }


\date{\today}

\begin{abstract}
The relativistic point-form formalism which we proposed for the study of the electroweak structure of few-body bound states is applied to calculate the elastic form factors of spin-1 mesons, such as the $\rho$, within constituent-quark models. We treat electron-meson scattering as a Poincar\'{e}-invariant coupled-channel problem for a Bakamjian-Thomas mass operator and extract the meson current from the resulting invariant 1-photon-exchange amplitude. Wrong cluster properties inherent in the Bakamjian-Thomas framework are seen to cause spurious contributions in the current. These contributions, however, can be separated unambiguously from the physical ones and we end up with a meson current with all required properties.
Numerical results for the $\rho$-meson form factors are presented assuming a simple harmonic-oscillator bound-state wave function.  The comparison with other approaches reveals a remarkable agreement  of our results with those obtained within the covariant light-front scheme proposed by Carbonell et al.

\end{abstract}


\maketitle


\section{\label{sec:intro} Introduction}
 Any proper description of a relativistic quantum-mechanical system consisting of interacting particles (or subsystems) should comply with Poincar\'{e} invariance. Another essential requirement is the property of cluster separability. It applies to systems in which subsystems can be isolated, which then should behave independently of the other subsystems. Both these physical principles not only demand certain properties from the Poincar\'{e} generators of the system, they also constrain the electromagnetic current operator describing the interaction of the system with an external field. Specifically, a correct current must transform as a four-vector operator under the Poincar\'{e} group. Further, cluster separability requires that the current must become the sum of the subsystem currents if all interactions between the subsystems are turned off. This is also related to the constraint that the charge of the whole system should be the sum of the subsystem charges, irrespective of whether the interaction is
present or not (for a detailed and formal discussion of these conditions, see the work by Lev, Ref.~\cite{Lev:1994au}). The current should also be conserved. Satisfying all these requirements makes the construction of a current for an interacting few-body system a non-trivial problem, since a bound state current must depend, in one or the other way, on the interaction between its constituents. The main purpose of the present work is to show for spin-1 two-body bound states, in particular the $\rho$-meson within the framework of constituent quark models, that such a current can actually be derived from a Poincar\'{e}-invariant coupled-channel approach to electron-meson scattering.

A particularly simple procedure for setting up a Poincar\'{e}-invariant framework for a quantum-mechanical system consisting of a finite number of interacting particles is the \textit{Bakamjian-Thomas} (BT) construction~\cite{Bakamjian:1953kh}. The central dynamical quantity in the BT framework is an invariant mass operator, from which the dynamical Poincar\'{e} generators follow. One advantage of the BT construction is that it still allows for instantaneous interactions, like in nonrelativistic quantum mechanics, without destroying Poincar\'{e} invariance. Another favorable feature is its natural connection to Dirac's forms of relativistic Hamiltonian dynamics~\cite{Dirac:1949cp}; the instant, the front and the point form. These stand for the three most simple, and yet the Poincar\'{e} algebra preserving ways of how to include interactions into a relativistic theory. \textit{Instant}, \textit{front} and \textit{point} characterize three different hypersurfaces in Minkowski space that are left invariant 
under the action of corresponding sets of Poincar\'{e} transformations that are not affected by interactions. These transformations, together with their generators are sometimes termed as \textit{kinematic}, whereas the remaining, interaction-dependent transformations and generators are rather called  \textit{dynamic}.

It has been proved by Sokolov and Shatnyi~\cite{Sokolov:1977im} that the three forms of relativistic dynamics are actually S-matrix equivalent, and therefore physically equivalent. However, they still differ strongly in their Poincar\'{e} transformation properties of operators and states. In the point form\footnote{For a short review on the point form we refer to Ref.~\cite{Biernat:2010tp}.} the transformation behaviour of states under Lorentz boosts and rotations is relatively simple due to the kinematic nature of the Lorentz group that is characteristic for this form. This also results in simple addition rules for angular momenta~\cite{Klink:1998zz}. When quantum systems with a finite number of degrees of freedom are treated within a point-form BT-framework, one commonly speaks of \textit{point-form relativistic quantum mechanics}. This constitutes the theoretical framework we adopt in this paper.

Point-form relativistic quantum mechanics has already been used previously to analyze the electromagnetic structure of simple hadronic few-body systems; see, for example, Refs.~\cite{Klink:1998qf,Allen:1998hb,Allen:2000ge,Wagenbrunn:2000es,Boffi:2001zb,Melde:2007zz}.
These papers employ the point-form spectator model to construct an electromagnetic current operator that satisfies all the requirements such as Poincar\'{e} covariance, current conservation and cluster separability.

In the present work we also use the point form to study the electromagnetic properties of vector mesons. We go, however, beyond making just an ansatz for the most general current on which the necessary constraints are imposed, and rather derive a microscopic meson current compatible with a particular interaction model, that exhibits the required properties. Our approach is based on the relativistic multi-channel framework proposed by Klink~\cite{Klink:2000pp}, with field-theoretical vertex interactions that are appropriately adapted to fit into the BT construction. By applying this framework to electron-meson scattering we have developed a general formalism to calculate electroweak meson currents. We have, for instance, already successfully calculated the electromagnetic form factor of the pion in Ref.~\cite{Biernat:2009my} and electroweak form factors of heavy-light systems in Ref.~\cite{GomezRocha:2012zd}.

As a next step in this program we focus in the present paper on the electromagnetic structure of vector mesons, such as the $\rho$-meson. The spin-1 case makes it necessary to carefully address the cluster problem. It is known that in a BT framework involving more than 2 particles one loses the property of cluster separability~(for a detailed discussion of this problem see Refs.~\cite{Sokolov:1977ym,Coester:1982vt,Keister:1991sb}). Within our formalism, this violation of cluster separability manifests itself in the appearance of additional structures in the meson current and additional dependencies in the meson form factors. Such additional structures and dependencies do not only come from the bound quark-antiquark system, but also from the the scattering  electron.
Although these additional contributions and dependencies show up already in the simple pion current, they are quite easily removed to obtain the pion form factor, see Ref.~\cite{Biernat:2009my}. This is, however, not the case for the more complex $\rho$-meson, where the proper extraction of the $\rho$-meson form factors requires a careful analysis of the current structure. Thereby, as will be shown in this paper, we find some quite remarkable similarities between our point-form approach and the covariant light-front approach of Refs.~\cite{Karmanov:1994ck,Carbonell:1998rj}, an insight that was not yet evident to us in our original work on the pion.

Formally it is known how to overcome the difficulties associated with cluster separability {within the BT framework}. For the three-particle case a solution to the cluster problem, formulated in terms of S-operators, has been given by Coester~in Ref.~\cite{Coester:1965zz}.  A general solution for an arbitrary number of particles has been proposed by Sokolov by introducing unitary operators, the \textit{Sokolov operators}, that restore cluster separability~\cite{Sokolov:1977ym}. In a recent work~\cite{PhysRevC.86.014002} Keister and Polyzou have tested -- for the first time and using a simple model -- to which quantitative extent the BT approach violates cluster separability. They have estimated the
corrections from the Sokolov construction needed to restore cluster separability. These estimates suggest that such corrections are too small for (weakly bound) nuclear systems to affect calculations of observables.

Although a Sokolov construction constitutes the proper solution to the cluster problem in the BT framework, it is rather formal and cumbersome for practical purposes. Therefore we have chosen an alternative way out. The idea is to identify the effects of wrong cluster properties that manifest themselves in the electromagnetic currents, and remove them in the sequel to end up with a unique physical current that has all required properties. Whether our procedure yields the identical results for the form factors as a proper Sokolov-corrected version of the BT approach is an interesting question that remains unanswered until the Sokolov corrections have been calculated explicitly for our model. This seems to be a quite intricate task. The size of the unphysical contributions in our current, however, will give us a good measure for the violation of cluster
separability for strongly bound systems like confined quark-antiquark pairs.

This paper is structured as follows. Section~\ref{sec:1} is devoted to a brief review of our coupled-channel point-form formalism for the derivation of electromagnetic meson currents. In Sec.~\ref{sec:pion} the pion current is reexamined. This also serves as a preparation for Sec.~\ref{sec:rho} where we derive and investigate the structure of the $\rho$-meson current. In Sec.~\ref{sec:numericalresults} the numerical results for the $\rho$-meson form factors are presented.  Section~\ref{sec:summary} contains the summary and an outlook.

\section{Meson current from electron-meson scattering}\label{sec:1}
We summarize briefly how the invariant 1-photon exchange amplitude and the electromagnetic meson current is derived within our point-form approach. The calculation is lengthy and tedious, and has already been given in detail in previous work~\cite{Biernat:2009my,Biernat:2011mp,GomezRocha:2012zd,Gomez-Rocha:2013bga}, hence we will restrict ourselves to just sketching it here.
\subsection{Optical potential}
We use the point-form formulation of the BT construction for a Poincar\'{e}-invariant treatment of interacting quantum-mechanical systems with a finite number of particles. In this framework the total four-momentum operator $\hat P^\mu$ of the interacting system is obtained from the product of an interacting mass operator $\hat {\mathcal M}$ and a free four-velocity operator  $\hat V_{\rm free}^\mu$,
\begin{equation}\label{eq:Pmu}
 \hat P^\mu=\hat{\mathcal M}\, \hat V_{\rm free}^\mu=(\hat{\mathcal M}_{\rm free}+\hat{\mathcal M}_{\rm int})\, \hat V_{\rm free}^\mu\,,
\end{equation}
where  $\hat{\mathcal M}_{\rm free}$ is the free mass operator and $\hat{\mathcal M}_{\rm int}$ is an interaction part that transforms like a Lorentz scalar and that commutes with $\hat V_{\rm free}^\mu$ to ensure Poincar\'{e} invariance. $\hat{\mathcal M}$ contains all information about the dynamics of the system and thus, by separating the overall motion of the system associated with $\hat V_{\rm free}^\mu$, the eigenvalue problem for $\hat P^\mu$ is reduced to an eigenvalue problem for the internal motion associated with $\hat{\mathcal M}$,
\begin{equation}\hat{\mathcal M}\, \vert \psi\rangle=m\, \vert \psi\rangle\,,
\label{eq:eigenvalueM}
\end{equation}
where $\vert \psi\rangle$ is the mass eigenstate of the system under consideration.
Since we want to account for the dynamics of the exchanged photon, we treat electron scattering off a confined quark-antiquark pair (meson) as a two-channel problem. The mass operator $\hat{\mathcal{M}}$ acts then on a Hilbert space that is the direct sum of $eq\bar q$ and $eq\bar q\gamma$ Hilbert spaces, where $e$, $q$, $\bar q$ and  $\gamma$ stand for electron, quark, antiquark and photon, respectively.\footnote{$q$ and $\bar q$ will sometimes be
referred to collectively as \lq\lq quarks''.}  As a consequence Eq.~(\ref{eq:eigenvalueM}) becomes a system of 2 coupled equations for $\vert \psi_{eq\bar q }\rangle$, the $eq\bar q$ component, and $\vert \psi_{eq\bar q\gamma }\rangle$, the $eq\bar q\gamma$ component
of $\vert \psi\rangle$.  After a Feshbach reduction the equation for $\vert \psi_{eq\bar q }\rangle$ reads~\cite{Klink:2000pp}
\begin{equation}
\label{eq:DynamicalEquationM} \left[\hat{M}_{eq\bar q}^{\rm conf}+\hat V_{\rm opt} (m)\right] \vert \psi_{eq\bar q }\rangle=m\vert \psi_{eq\bar q }\rangle\, ,
\end{equation}
where
\begin{eqnarray} \label{eq:Vopt}
 \hat V_{\rm opt} (m)=
\hat{K}_\gamma \left(m-\hat{M}_{eq\bar q\gamma}^{\rm conf}
\right)^{-1}\hat{K}_\gamma^\dag
\end{eqnarray}
is the optical potential. Here $\hat{K}_\gamma^{(\dag)}$ is a vertex operator that describes the absorption (emission) of the photon by the electron, quark or antiquark. $\hat{M}_{eq\bar q}^{\rm conf}$ and $\hat{M}_{eq\bar q\gamma}^{\rm conf}$ are the invariant mass operators of the $eq\bar q$ and $eq\bar q\gamma$  systems, respectively. They include an instantaneous confining interaction between $q$ and $\bar q$. For instance, $\hat{M}_{eq\bar q}^{\rm conf}$ is defined by
\begin{equation}\hat{M}_{eq\bar q}^{\rm conf}=\hat{M}_{eq\bar q}+\hat V_{eq\bar q}^{\rm conf} \label{eq:Mint}\, ,
\end{equation}
where $\hat{M}_{eq\bar q}$ is the mass operator of the free $eq\bar q$ system and $\hat V_{eq\bar q}^{\rm conf}$ denotes the embedding of the confining $q\bar q$ potential in the $eq\bar q$ Hilbert space. $\hat{M}_{eq\bar q\gamma}^{\rm conf}$ is defined analogously. The optical potential $\hat{V}_{\mathrm{opt}}(m)$ consists of all possible exchanges of the photon between the electron and the quarks including loop contributions, i.e. reabsorption by the emitting particle. The factor $(\hat{M}_{eq\bar q\gamma}^{\rm conf}-m)^{-1}$ in Eq.~(\ref{eq:Vopt}) describes the propagation of the $e q\bar q \gamma$ intermediate state and is thus  responsible for retardation effects.

\subsection{Meson current}

The electromagnetic meson current can be extracted from the elastic electron-meson scattering amplitude calculated in the 1-photon-exchange approximation. We do this in our point-form BT formulation where the 1-photon-exchange amplitude is obtained from appropriate matrix elements of the optical potential~(\ref{eq:Vopt}) between, so-called, velocity states for the electron and the confined $q\bar q$ system. The basis of velocity states~\cite{Klink:1998zz} is a natural basis for multiparticle states in the point-form BT framework, as the overall four-velocity is not affected upon introducing interactions, see Eq.~(\ref{eq:Pmu}). An $n$-particle velocity state, denoted by $\vert
v; \vec{k}_1, \mu_1;\dots; \vec{k}_n, \mu_n
\rangle $, is an $n$-particle momentum state with rest-frame momenta $\vec{k}_1,\ldots ,\vec{k}_n$ (satisfying $\sum_{i=1}^n \vec{k}_i=0$) and spin projections $\mu_1,\ldots, \mu_n$ that is boosted to overall four-velocity $v$ (with $v_\mu v^\mu = 1$) by means of a canonical spin boost $B_c(v)$~\cite{Keister:1991sb}. Velocity states form a complete orthogonal basis, they are eigenstates of the invariant $n$-particle (free) mass operator and they have a rather simple behavior under Lorentz transformations as compared to the usual momentum states (details on the properties of velocity states can be found, for instance, in Refs.~\cite{Klink:1998zz,Krassnigg:2003gh}).

The relevant matrix elements of the optical potential, from which the meson current is extracted, are
\begin{eqnarray}\label{eq:voptclust}
\langle  \underline{v}^\prime;
\vec{\underline{k}}_e^\prime, \underline{\mu}_e^\prime;
\underline{\vec{k}}_\alpha^\prime,\underline{\mu}_\alpha^\prime,
\alpha \vert\,
\hat{V}_{\mathrm{opt}}(m)
\vert\, \underline{v}; \vec{\underline{k}}_e, \underline{\mu}_e;
\underline{\vec{k}}_\alpha,\underline{\mu}_\alpha,
\alpha \rangle_{\mathrm{os}}\, .
\end{eqnarray}
Here $\alpha$ is a shorthand notation for the discrete quantum numbers necessary to uniquely specify the meson of interest.
$\underline{v}^{(\prime)}$ is the incoming (outgoing) overall four-velocity of the electron-meson system,
$\vec{\underline{k}}_e^{(\prime)}$, $\underline{\vec{k}}_\alpha^{(\prime)}$ and $\underline{\mu}_e^{(\prime)}$, $\underline{\mu}_\alpha^{(\prime)}$ are the momenta and spin projections of the incoming (outgoing) electron and meson as defined in the electron-meson rest frame.  Here we have introduced the underlining of velocities, spins and momenta for states where the quark and the antiquark are confined (forming the meson) to make a clear distinction from states where the quark and the antiquark are free particles. Since we consider elastic electron-meson scattering we can restrict our considerations to \lq on-shell' matrix elements
[denoted by the \lq\lq os\rq\rq\ subscript in Eq.~(\ref{eq:voptclust})], for which the total invariant mass of incoming and outgoing electron and meson is the same, i.e.
\begin{eqnarray}\label{eq:m}
 m&= & \sqrt{s}= \omega_{\underline{k}_e}+\omega_{\underline{k}_\alpha} =
\omega_{\underline{k}_e^\prime}+\omega_{\underline{k}_\alpha^\prime}\, ,\\
\omega_{\underline{k}_e}&=&\omega_{\underline{k}_e^\prime}\,, \quad \label{eq:meqmp2}
\omega_{\underline{k}_\alpha}=\omega_{\underline{k}_\alpha^\prime}\,,
\end{eqnarray}
where $\omega_{\underline{k}_i}=\sqrt{\vec{\underline{k}}_i^2+m_i^2}$ with $i=e,\,\alpha$.

The first step of evaluating the matrix elements~(\ref{eq:voptclust}) of the optical potential~(\ref{eq:Vopt}) is a multiple insertion of completeness relations for velocity eigenstates of $\hat{M}_{eq\bar q(\gamma)}^{\rm conf}$ and $\hat{M}_{eq\bar q(\gamma)}$ at the appropriate places, which gives rise to velocity-state matrix elements of the form
\begin{eqnarray}\label{eq:wfmatrixele}
 \langle v; \vec{k}_e, \mu_e; \vec{k}_q, \mu_q; \vec{k}_{\bar{q}},
\mu_{\bar{q}} \vert\, \underline{v}; \vec{\underline{k}}_e,
\underline{\mu}_e;
\underline{\vec{k}}_\alpha,\underline{\mu}_\alpha, \alpha
\rangle\,,
\end{eqnarray}
\begin{eqnarray}
\langle v; \vec{k}_e, \mu_e; \vec{k}_q, \mu_q; \vec{k}_{\bar{q}},
\mu_{\bar{q}};\vec{k}_{\gamma}, \mu_{\gamma} \vert\,
\underline{v}; \vec{\underline{k}}_e, \underline{\mu}_e;
\underline{\vec{k}}_\alpha,\underline{\mu}_\alpha, \alpha;
\vec{\underline{k}}_\gamma, \underline{\mu}_\gamma\rangle\,\nonumber\\
\end{eqnarray}
and
\begin{eqnarray}\label{eq:Kvertex}
 \langle v^\prime; \vec{k}_e^\prime,\! \mu_e^\prime;
\vec{k}_q^\prime,\! \mu_q^\prime; \vec{k}_{\bar{q}}^\prime,\!
\mu_{\bar{q}}^\prime; \vec{k}_\gamma^\prime,\! \mu_\gamma^\prime
\vert \,\hat{K}^\dag\, \vert v\,  ; \vec{k}_e,\! \mu_e; \vec{k}_q,\!
\mu_q; \vec{k}_{\bar{q}},\! \mu_{\bar{q}} \rangle\,,\nonumber\\
\end{eqnarray}
together with their Hermitian conjugates, respectively. The first two expressions are proportional to the wave function $\psi_{\alpha\underline{\mu}_\alpha\mu_q\mu_{\bar{q}}}(\vec{\tilde k}_q)$ of the confined $q\bar{q}$ pair (meson). The tilde refers to the rest frame of the $q\bar q$ subsystem, i.e. $\tilde k_i= B_{\rm c}^{-1}(v_{q\bar q})k_i$ with $i=q,\,\bar q$ where $v_{q\bar q}=(k_q+k_{\bar q})/m_{q\bar q}$ is the four-velocity of the free $q\bar q$ pair in the overall rest frame and
\begin{eqnarray}
 m_{q\bar q}=\omega_{\tilde k_q}+\omega_{\tilde k_{\bar q}}=\sqrt{\left(\omega_{k_q}+\omega_{k_{\bar q}}\right)^2-\left(\vec k_q+\vec k_{\bar q }\right)^2}
\end{eqnarray}
the invariant mass of the free $q \bar q$ pair. Note that the center-of-mass kinematics associated with the velocity states implies $\underline{\vec{k}}_\alpha=\vec k_q+\vec k_{\bar q}$ and therefore $\underline{\vec{k}}_\alpha/m_{q\bar q}\equiv \vec v_{q\bar q}$.

The third expression, Eq.~(\ref{eq:Kvertex}), describes the transition from the free
$eq\bar{q}$ state to the free $eq\bar{q}\gamma$ state by emission of a
photon. It is calculated from the usual field-theoretical interaction density
${\mathcal{L}}^{\mathrm{em}}_{\mathrm{int}}(x)$ of spinor quantum
electrodynamics which involves the (conserved) pointlike current operators of the quarks and the electron~\cite{Klink:2000pp}. Explicit formulae for all matrix elements~(\ref{eq:wfmatrixele})-(\ref{eq:Kvertex}) together with their Hermitian conjugates can be found in Refs.~\cite{Biernat:2009my,Biernat:2011mp}. The necessary integrations and sums from the multiple insertion of the completeness relations in (\ref{eq:voptclust}) can be done by means of the appropriate Dirac and Kronecker deltas, respectively. Neglecting the 3 contributions where the photon is reabsorbed by the emitting particle and another 2 where the photon is exchanged between quark and antiquark, as these are just (electromagnetic) self-energy corrections of electron and meson masses\footnote{ Due to instantaneous confinement mass renormalization happens on hadron rather than on quark level.}, the remaining 4 time-ordered contributions can be combined to 2 covariant contributions that correspond to photon exchange between electron and 
either quark or antiquark. The final result for the invariant 1-photon-exchange amplitude, as given by Eq.~(\ref{eq:voptclust}), has the expected structure. It is a contraction of the (pointlike) electron current $-e\,\bar{u}_{\underline{\mu}_e^\prime}
(\vec{\underline{k}}_e^\prime)\gamma^\mu u_{\underline{\mu}_e}
(\vec{\underline{k}}_e)$ with the meson current $eJ_\alpha^\nu$ (that contains the $q\bar{q}$ bound-state wave function and the (anti)quark current) multiplied with the covariant photon propagator $(-\mathrm g_{\mu \nu})/Q^2$~\cite{Biernat:2009my,Biernat:2011mp}:
\begin{widetext}
 \begin{eqnarray}\label{eq:voptos}
\lefteqn{\langle  \underline{v}^\prime;
\vec{\underline{k}}_e^\prime, \underline{\mu}_e^\prime;
\underline{\vec{k}}_\alpha^\prime,\underline{\mu}_\alpha^\prime,
\alpha \vert\,
\hat{V}_{\mathrm{opt}}(m)
\vert\, \underline{v}; \vec{\underline{k}}_e, \underline{\mu}_e;
\underline{\vec{k}}_\alpha,\underline{\mu}_\alpha, \alpha \rangle_{\mathrm{os}}}
\nonumber \\
&&=\underline{v}_0 \delta^3 (\vec{\underline{v}}^{\, \prime} -
\vec{\underline{v}}\, )\, \frac{(2 \pi)^3
}{\sqrt{(\omega_{\underline{k}_e^{\prime}}+
\omega_{\underline{k}_\alpha^{\prime}})^3}
\sqrt{(\omega_{\underline{k}_e^{\phantom{\prime}}}+
\omega_{\underline{k}_\alpha^{\phantom{\prime}}})^3}}
(-e^2)\,\bar{u}_{\underline{\mu}_e^\prime}
(\vec{\underline{k}}_e^\prime)\gamma^\mu u_{\underline{\mu}_e}
(\vec{\underline{k}}_e)\frac{(-\mathrm g_{\mu \nu})}{Q^2}J_\alpha^\nu(\underline{\vec{k}}_\alpha^\prime,
\underline{\mu}_\alpha^\prime;\underline{\vec{k}}_\alpha,
\underline{\mu}_\alpha)\, .
\end{eqnarray}
The denominator of the photon propagator is given by $Q^2=-\underline{q}_\mu \underline{q}^\mu$, with $\underline{q}^\mu = (\underline{k}_\alpha^\prime-
\underline{k}_\alpha)^\mu$ denoting the four-momentum transferred between electron and meson.
The meson current reads
 \begin{eqnarray}
J_\alpha^\nu(\underline{\vec{k}}_\alpha^\prime,
\underline{\mu}_\alpha^\prime;\underline{\vec{k}}_\alpha,
\underline{\mu}_\alpha)&=&
\sqrt{\omega_{\underline{k}_\alpha}\omega_{\underline{k}_\alpha^{\prime}}}\sum_{\mu_q'\mu_{\bar q}'
} \bigg[\int \frac{\mathrm d^3 k_{q}'}{\omega_{k_q'}}\frac{1}{\omega_{k_{\bar q}'}}\frac{1}{ \omega_{k_q}}\sqrt{\frac{\omega_{\tilde k_q'} \omega_{\tilde k_{\bar q}'}}{\omega_{\tilde k_q'}+\omega_{\tilde k_{\bar q}'}}}\sqrt{\frac{\omega_{\tilde k_q} \omega_{\tilde k_{\bar q}}}{\omega_{\tilde k_q}+\omega_{\tilde k_{\bar q}}}} \sqrt{ \omega_{k_{q}'}+\omega_{k_{\bar q}'}}\sqrt{\omega_{k_{q}}+\omega_{k_{\bar q}}} \nonumber\\&&\times\sum_{\mu_q} \psi^\ast_{\alpha\underline{\mu}'_\alpha\mu_q'\mu_{\bar{q}}'}(\vec{\tilde k}_q') \, \psi_{\alpha\underline{\mu}_\alpha\mu_q\mu_{\bar{q}}'}(\vec{\tilde k}_q)
 \, Q_q\,  j_q^\nu(\vec{k}_q^\prime,\mu_q^\prime;\vec{k}_q,\mu_q)+ (q\leftrightarrow \bar q) \bigg] \,\, ,  \label{eq:Jalpha}
 \end{eqnarray}
 with $Q_q$ denoting the charge of the quark in units of $|e|$ and the (pointlike) currents for quark and antiquark being defined as
 \begin{eqnarray}
  j_q^\nu( \vec{k}_q^\prime,\mu_q^\prime;\vec{k}_q,\mu_q)=\bar{u}_{\mu_q^\prime}(\vec{k}_q^\prime)\,\gamma^\nu\,
u_{\mu_q}(\vec{k}_q) \quad\text{and} \quad
j_{\bar q}^\nu(\vec{k}_{\bar q}^\prime,\mu_{\bar q}^\prime; \vec{k}_{\bar q},\mu_{\bar q})=\bar{v}_{\mu_{\bar q} }(\vec{k}_{\bar q} )\,\gamma^\nu\,
v_{\mu_{\bar q}^\prime }(\vec{k}_{\bar q}^\prime )\, ,\label{eq:jq}
 \end{eqnarray}
 respectively. The meson wave function is given by
 \begin{eqnarray}
 &&\psi_{\alpha\underline{\mu}_\alpha\mu_q\mu_{\bar{q}}}(\vec{\tilde k}_q)\equiv \psi _{nj_\alpha\underline{\mu}_\alpha\mu_q\mu_{\bar{q}}}(\vec {\tilde k}_q)\nonumber\\&&:=
\sum_{ls\mu_l\mu_s\tilde \mu_q\tilde \mu_{\bar q}}
Y_{l\mu_l}\left(\sfrac{\vec{\tilde{ k}}_q}{|\vec{\tilde{k}}_q|}\right)
C^{s\mu_s}_{\frac12\tilde \mu_q\frac12\tilde \mu_{\bar q}}C^{j_\alpha\underline{\mu}_\alpha}_{l\mu_ls\mu_s} u_{nls}^{j_{\alpha}}(|\vec{\tilde{k}}_q|)
 D^{\frac12}_{\mu_q\tilde \mu_q}\left[ R_W\left(\sfrac{\tilde k_q}{m_q} ,B_{c}(v_{q\bar q})\right)\right]
D^{\frac12}_{\mu_{\bar q}\tilde \mu_{\bar q}}\left[ R_W\left(\sfrac{\tilde k_{\bar q}}{m_{\bar q}}, B_{c}( v_{q\bar q})\right)\right]\,  \label{eq:mesonwf}
 \end{eqnarray}
\end{widetext}
 where $n,j_\alpha, l$ and $s$ are the quantum numbers of radial excitations, total angular momentum, orbital angular momentum and total spin, respectively, with $\underline{\mu}_\alpha, \mu_l$ and $\mu_s$ the corresponding projections on the $z$-axis. $Y_{l\mu_l}(\vec{\tilde{ k}}_q/|\vec{\tilde{k}}_q|)$ and $u_{nls}^{j_{\alpha}}(|\vec{\tilde{k}}_q|)$ are the usual spherical harmonics and the radial wave functions, respectively. $C^{j_\alpha\underline{\mu}_\alpha}_{l\mu_ls\mu_s}$ are the usual Clebsch-Gordan coefficients. $D^{\frac12}_{\mu_i\tilde \mu_i}\left[ R_W(\ldots)\right]$ are the Wigner D-functions where $R_W(\ldots)=R_W\left(\tilde k_i/m_i ,B_{c}(v_{q\bar q})\right)$ with $i=q,\, \bar q$ is the Wigner rotation (associated with canonical-spin boosts)
  \begin{eqnarray}
   R_W\left(\sfrac{\tilde k_i}{m_i} ,B_{c}(v_{q\bar q})\right)= B_{c}^{-1}\left(\sfrac{k_i}{m_i}\right)B_{c}(v_{q\bar q})B_{c}\left(\sfrac{\tilde k_i}{m_i}\right)\,.\nonumber\\
  \end{eqnarray}
The wave function is normalized to unity:
\begin{eqnarray}\label{eq:normeigenfucntions1}
&&\int \mathrm d^3 \tilde k_q\sum_{\mu_q\mu_{\bar q}} \psi^\ast _{nj_\alpha\underline{\mu}_\alpha\mu_q\mu_{\bar q}}(\tilde{\vec k}_q)\psi _{n'j_\alpha'\underline{\mu}_\alpha'\mu_q\mu_{\bar q}}(\tilde{\vec k}_q)
 \nonumber\\&&\quad=\delta_{nn'}\delta_{j_\alpha j_\alpha'}\delta_{\underline{\mu}_\alpha\underline{\mu}_\alpha'}\,.
\end{eqnarray}
In Eq.~(\ref{eq:Jalpha}) the quark momenta  with and without prime are related by
  $\vec{k}_i^{ \prime} =
\vec{k}_i + \vec{q}
= \vec{k}_i + \vec{k}_\gamma$ where $i$ denotes the active quark (note that the inactive quark must satisfy spectator conditions). This means three-momentum conservation at the electromagnetic vertices, a property which one would not expect
in point-form quantum mechanics. One should, however, keep in mind that we are dealing with overall-center-of-mass momenta when working with the velocity-state representation and the energy is not conserved at the vertices. For the physical momenta, i.e. the center-of-mass momenta boosted by $B_c(\underline v)$, none of the four-momentum
components is, in general (if $\vec {\underline v} \neq 0$), conserved at the electromagnetic vertices. It should also be mentioned that, in general, the four-momentum transfer between incoming and outgoing (active) quark $q^\mu:=(k_q^\prime-k_q)^\mu$ deviates from the four-momentum transfer between incoming and outgoing confined $q\bar q$ pair $\underline q^\mu$. While the three-momentum transfers are the same, i.e. $\vec{q}=\vec{\underline q}$ due to the center-of-mass kinematics, the zero components differ, $q^0\neq \underline q^0=0$, because of~(\ref{eq:meqmp2}) and
$\omega_{k_q^\prime}\neq \omega_{k_q}$. Therefore, not all the four-momentum that is transferred via the photon to the $q\bar q$ bound state is also transferred to the active quark.

 In the present work we restrict ourselves to $q\bar q$-mesons consisting of quark and antiquark with equal masses $m_q=m_{\bar q}$. In this case the quark and antiquark currents~(\ref{eq:jq}) are identical, i.e. $j_q^\nu( \vec{k}_q^\prime,\mu_q^\prime;\vec{k}_q,\mu_q)=j_{\bar q}^\nu(\vec{k}_{\bar q}^\prime,\mu_{\bar q}^\prime; \vec{k}_{\bar q},\mu_{\bar q})$. For the treatment of systems with unequal quark masses within the present formalism, like heavy-light mesons, we refer to Refs.~\cite{GomezRocha:2012zd,Gomez-Rocha:2013bga}.
\section{Pion}\label{sec:pion}
In this section we consider the case of the $q\bar q$ bound state being a charged pseudoscalar meson with total angular momentum $j_\alpha=0$, such as the pion. The pion has already been studied in this framework in Ref.~\cite{Biernat:2009my}. The reason why we review the $j_\alpha=0$  case is to prepare the reader for the more complex, but in some aspects similar $j_\alpha=1$ case of charged vector mesons.

A positively (negatively) charged $\pi^{+} (\pi^-)$ meson is described in the constituent-quark model as a confined $u \bar d$ ($\bar u d$) pair. Assuming equal $u$- and $d$-quark masses and a pure s-wave ($l=0$) the current~(\ref{eq:Jalpha}) simplifies for the case of a pion to~\cite{Biernat:2011mp}
\begin{widetext}
\begin{eqnarray}
 J_{\pi}^\mu(\underline{\vec{k}}_\alpha^\prime,\underline{\vec{k}}_\alpha)&=&\frac {\sqrt{\omega_{\underline{k}_\alpha}\omega_{\underline{k}_\alpha^{\prime}}}}{8\pi}
\int \frac{\mathrm d^3 \tilde k_{q}'}{\omega_{k_q}}\sqrt{\frac{m_{q\bar q}}{m_{q\bar q}'}}\sqrt{\frac{\omega_{k_q} +\omega_{k_{\bar q}}}{\omega_{k_q'}+\omega_{k_{\bar q}'}}} u^\ast_{n0}(|\vec{\tilde{k}}_q'|)u_{n0}(|\vec{\tilde{k}}_q|)\nonumber\\&&\times\sum_{\mu_q\mu_q'}D^{\frac12}_{\mu_q\mu_q'}\left[ R_W\left(\sfrac{\tilde k_q}{m_q}, B_{c}(v_{q\bar q})\right)
R_W^{-1}\left(\sfrac{\tilde k_{\bar q}}{m_q}, B^{-1}_{c}(v_{q\bar q}') B_{c}(v_{q\bar q})\right)
R_W^{-1}\left(\sfrac{\tilde k_q'}{m_q}, B_{c}(v_{q\bar q}')\right)\right]
 \nonumber\\&&\times(Q_q+Q_{\bar q}) j_q^\mu(\vec{k}_q^\prime,\mu_q^\prime;\vec{k}_q,\mu_q) \,.\nonumber\\  \label{eq:Jpi}
\end{eqnarray}
\end{widetext}
In Ref.~\cite{Biernat:2011mp} fundamental properties of the current, like hermiticity, covariance and continuity, have been investigated in some detail (the corresponding proofs are rather lengthy and can be found in Sec. 4.4 and App. D of Ref.~\cite{Biernat:2011mp}):
(i) The pion current satisfies the property of hermiticity, i.e. $[J_{\pi}^{\mu}(\underline{\vec{k}}_\alpha^\prime,
\underline{\vec{k}}_\alpha)]^\dag=J_{\pi}^\mu(\underline{\vec{k}}_\alpha,
\underline{\vec{k}}_\alpha^\prime)$.\\
(ii) The correct behavior under Lorentz transformations is guaranteed by the current
\begin{eqnarray}
J_{\pi}^\mu(\vec{p}_\alpha^\prime,\vec{p}_\alpha):= [B_{c}(\underline v)]^\mu_{\,\,\nu}J_{\pi}^\nu(\underline{\vec{k}}_\alpha^\prime,\underline{\vec{k}}_\alpha)\label{eq:picurrent}
\end{eqnarray}
that depends on the \emph{physical} pion momenta
$p_\alpha^{(\prime)}=B_{c}(\underline v)\underline{k}_\alpha^{(\prime)} $ instead of $\underline{k}_\alpha^{(\prime)}$, the momenta in the electron-meson rest frame that originate from the velocity-state representation.\\
(iii) The pion current is conserved, i.e. \\ \phantom{a} \hspace{1cm} $(p_\alpha'-p_\alpha)_\mu J_{\pi}^\mu(\vec{p}_\alpha^\prime,\vec{p}_\alpha)=0$.
\subsection{Covariant structure of the current}
The correct physical pion current, denoted by $I^\mu_{\pi}(\vec p_\alpha^\prime,\vec p_\alpha)$, can be expressed in terms of only one covariant, the sum of incoming and outgoing  physical pion 4-momenta, $P_\alpha=p_\alpha+p_\alpha'$. This covariant is multiplied by the electromagnetic pion form factor $F$, with $F$ being a function of Mandelstam $t=-Q^2$. Correct cluster properties in this context mean that the current cannot depend on the presence of other particles, like the projectile. It turns out, however, that this is not the case for our electromagnetic pion current given by Eqs.~(\ref{eq:picurrent}) and (\ref{eq:Jpi}), as it exhibits an additional dependence on the incoming and outgoing electron momenta  $\underline p_e$ and $\underline p_e'$. The reason becomes clear by the following analysis: For spinless particles, like the pion, we get 4 current components $J^\mu_{\pi}(\vec p_\alpha^\prime,\vec p_\alpha)$, $\mu=0,1,2,3$.
Due to rotational invariance of our approach the scattering plane can be chosen such that one of the (space) components vanishes. As a consequence of current  conservation only 2 of the remaining 3 non-vanishing current components can be independent. That there are indeed 2 independent current components is revealed by the numerical analysis~\cite{Biernat:2009my,Biernat:2011mp}. A covariant decomposition of $J^\mu_{\pi}(\vec p_\alpha^\prime,\vec p_\alpha)$ is thus accomplished by means of two current-conserving four-vectors which are multiplied with corresponding form factors $f$ and $b$. The only current-conserving four-vector that can be built from the incoming and outgoing pion momenta is obviously $P_\alpha^\mu$, the sum
of both momenta. Looking for a second covariant we have to recall that our derivation of the current is based on the Bakamjian-Thomas construction which is known to provide wrong cluster properties for more than 2 particles~\cite{Keister:1991sb}. This means that the physical properties of our model pion  may depend on the presence of an additional particle, such as the electron. It is thus quite tempting to choose as a second, current-conserving four-vector the sum of the incoming and outgoing electron momenta $P_e=\underline p_e+\underline p_e'$.

Wrong cluster properties do not only modify the covariant structure of our model current, they also affect the coefficients in front of the covariants, the form factors $f$ and $b$. These do not only depend on Mandelstam $t=-Q^2$, the four-momentum-transfer squared, but also on Mandelstam
\begin{eqnarray}\label{eq:Mandelstams}
  s=(p_\alpha+p_e)^2=\left(\sqrt{m_{\pi}^2+\underline{\vec k}_\alpha^2}+\sqrt{m_{e}^2+\underline{\vec k}_\alpha^2}\right)^2\,,
 \end{eqnarray}
the square of the invariant mass of the electron-pion system.
The $s$-dependence can equivalently be expressed as a dependence on the magnitude of the particle momenta (in the electron-pion rest frame)
\begin{eqnarray}\label{eq:kmagnitude}
 k&:=&|\underline{\vec k}_\alpha^{\prime}|=|\vec k_e^{\prime}|=|\underline{\vec k}_\alpha|=|\vec k_e|\, ,
\end{eqnarray}
where we have used Eq.~(\ref{eq:meqmp2}). Eq.~(\ref{eq:Mandelstams}) can be inverted to relate $k$ and $s$:
\begin{eqnarray}\label{eq:kmagnitude2}
k^2=\frac{(m_\pi^2-m_e^2)^2+s\left[s-2(m_\pi^2+ m_e^2)\right]}{4s}\,.
\end{eqnarray}
Using $k$ instead of $s$ turns out to be more convenient for our purposes.  At this point it should be mentioned that Poincar\'e invariance of our Bakamjian-Thomas type approach is not spoiled by vertex form factors that are functions of a whole set of independent Lorentz invariants involved in the process. However, a reasonable microscopic model for electromagnetic form factors should, of course, only depend on the momentum transfer squared $t$ and not on $s$. Fortunately, as discussed later, the unwanted $s$-dependence can be eliminated in a certain limit.

With these findings the general covariant decomposition of our pion current reads
 \begin{eqnarray}\label{eq:formffPS}
 J_{\pi}^\mu(\vec{p}_\alpha^\prime,\vec{p}_\alpha)=
\,f(Q^2,k)P_\alpha^\mu +b (Q^2,k)P_e^\mu\,.
\end{eqnarray}
This decomposition holds for arbitrary values of the pion momenta $p_\alpha$ and $p_\alpha^\prime$ with one exception, the so-called \textit{Breit frame} which corresponds to ${\vec p}_\alpha=-{\vec p}_\alpha'$ ($={-\vec p}_e={\vec p}_e'$). In this frame the two covariants $P_\alpha$ and $P_e$ become proportional which precludes the separation of the two form factors.

There seems to be an ambiguity how to define the form factors by expanding the current in terms of covariants. It turns out, however, that only the form factor $f(Q^2,k)$ defined via the expansion~(\ref{eq:formffPS}) provides the correct charge of the pion at $Q^2=0$, as it is required for the physical form factor. This justifies to call $f$ defined in Eq.~(\ref{eq:formffPS}) the \textit{physical form factor} of the pion. The remaining structure in Eq.~(\ref{eq:formffPS}) that is proportional to the sum of electron momenta will be referred to as \textit{non-physical} (or \textit{spurious}) contribution with $b$ being the \textit{spurious form factor}. Hence, only the expansion~(\ref{eq:formffPS}) provides a sensible separation of the physical from the spurious contribution. The separation of Eq.~(\ref{eq:formffPS}) suggests the following definition: spurious contributions to the current are defined as all structures that depend on the sum of electron momenta.

The covariant structure of our current resembles the corresponding one obtained in a covariant light-front approach~\cite{Karmanov:1994ck,Carbonell:1998rj}.
In these papers the authors encounter a spurious (unphysical) contribution to the current which is associated with the light-like four-vector $\omega^\mu$ that defines the orientation of the light front (defined by the equality $\omega\cdot x=0$). Their spurious contribution is comparable to our spurious contribution if $\omega^\mu$ is identified with $P_e^\mu$. Our spurious terms in the current can be traced back to the violation of cluster separability in our point-form Bakamjian-Thomas approach. In the covariant light-front formalism the spurious $\omega$-dependent contribution is rather the consequence of the most general ansatz for a pion current that has to include the orientation of the light front.

\subsection{Electromagnetic form factor}
The form factors are functions of the Lorentz invariants $t$ and $s$ and can therefore be extracted in any inertial frame. For simplicity we choose the electron-meson rest frame in which $\vec {\underline v} = 0$, and thus $p_\alpha^{(\prime)}=\underline{k}_\alpha^{(\prime)}$ with
\begin{equation}\label{eq:momentumscatt}
\underline{\vec{k}}_\alpha=-\underline{\vec{k}}_e=
\left(
\begin{array}{c} -\frac{Q}{2}\\ 0 \\
\sqrt{k^2-\frac{Q^2}{4}}
\end{array}
\right)\,  \quad \hbox{and}\quad \vec{q}=\left(
\begin{array}{c} Q\\ 0 \\ 0 \end{array}\right)\,.
\end{equation}
In this parametrization $k$ is subject to the constraint that $k\geq \sfrac Q2$. The only non-vanishing components of the pion current in this frame are $J_\pi^0$ and $J_\pi^3$ from which the form factors $f(Q^2,k)$ and $b(Q^2,k)$ can be extracted by inserting our microscopic expression for the pion current, Eq.~(\ref{eq:Jpi}),  into the left-hand side of Eq.~(\ref{eq:formffPS}).
A numerical analysis of the resulting $f(Q^2,k)$ and $b(Q^2,k)$, as presented in Ref.~\cite{Biernat:2011mp}, confirms that both, the physical form factor $f(Q^2,k)$ and the spurious form factor $b(Q^2,k)$ do not only depend on $Q^2$, but they depend indeed also on $k$. However, this $k$-dependence of $f(Q^2,k)$ vanishes rather quickly with increasing $k$ (or equivalently increasing Mandelstam $s$). At the same time the spurious form factor $b(Q^2,k)$ is seen to vanish.  It is thus suggestive to take the limit $k\rightarrow\infty$ to get a sensible result for the physical form factor $f(Q^2,k)$ that only depends on $Q^2$. As a further benefit of this limit one gets rid of the unwanted spurious contribution on the right-hand side of Eq.~(\ref{eq:formffPS}) since $b(Q^2,k)$ vanishes.
After analyzing the integrand on the right-hand side of Eq.~(\ref{eq:Jpi}) in the limit $k\rightarrow\infty$ (for details see Ref.~\cite{Biernat:2011mp}) we find for the pion current~\cite{Biernat:2009my}
\begin{eqnarray}
 I^\mu_\pi:= \lim_{k \rightarrow
\infty}J^\mu_{\pi}(\vec p_\alpha^\prime; \vec p_\alpha )= \,F(Q^2)\lim_{k \rightarrow
\infty} P_\alpha^\mu\, , \label{eq:formfactorF}
\end{eqnarray}
since $J^0_{\pi}\rightarrow J^3_{\pi}$ in this limit (in the electron-pion rest frame)
with the pion form factor
\begin{eqnarray}
 F(Q^2)&:=&\lim_{k \rightarrow
\infty} f(Q^2,k)\nonumber\\&=&
\frac{1}{4\pi}\int\mathrm{d}^3\tilde{k}^\prime_q
\sqrt{\sfrac{m_{q\bar q}}{m'_{q\bar q}}}\, \mathcal{S}\, u_{n
0}^\ast\,(|\vec {\tilde k}_q^\prime|)\, u_{n 0}\,(|\vec {\tilde k}_q|)\, , \nonumber\\ \label{eq:pionformfactor}
\end{eqnarray}
where the spin-rotation factor reads
\begin{eqnarray}
\mathcal{S}=\frac{m_{q\bar q}^\prime}{m_{q\bar q}}-\frac{2
\tilde{k}_q^{\prime 1} \, Q}{m_{q\bar q} (m_{q\bar q}^\prime + 2
\tilde{k}_q^{\prime 3})}\,.
\end{eqnarray}

In Ref.~\cite{Biernat:2009my} we found, after a simple change of integration variables, that the pion form factor result~(\ref{eq:pionformfactor}) is actually \emph{identical} to the result obtained in the  usual light-front approach of Refs.~\cite{Chung:1988mu,Simula:2002vm}. Therein, use is made of $q^+=q^0+q^3=0$ frames, which has the advantage that Z-graphs are suppressed. This remarkable equivalence between point-form and light-front approaches can be better understood by first noting that the  usual light-front approach corresponds to the special case of the covariant light-front approach where the orientation of the light front is fixed by $\omega=(1,0,0,-1)$. In the usual approach the pion form factor is then extracted from the plus
component $J_\pi^+$ of the current. Since $\omega^+=0$ for $\omega=(1,0,0,-1)$, the spurious part of the current, proportional to $\omega$, does not contribute to $J_\pi^+$ and therefore also not to the pion form factor extracted from $J_\pi^+$~\cite{Karmanov:1994ck}. Consequently, the
usual light-front and the covariant light-front dynamics give the same pion form factor, which is in some way an exception due to the simplicity of spin-0 systems and does not hold for the more complex spin-1 systems, as we will see soon.
In fact, taking the plus component of the current to extract the form factor in the usual light-front approach plays a similar role as the limit $k\rightarrow\infty$ (or equivalently $s\rightarrow\infty$) in our approach.  It removes the spurious contribution in the current.

It is also possible to project out the form factor directly from the current like it has been proposed in Refs.~\cite{Karmanov:1994ck,Carbonell:1998rj}. Contracting the pseudoscalar bound-state current with the four-vector $K^\mu_e/(K_\alpha\cdot K_e)$ in the limit $k \rightarrow
\infty$ gives the form factor:
 \begin{eqnarray}\label{eq:projoutF}
 F(Q^2)=\lim_{k \rightarrow
\infty}\frac{K_{e\mu}}{K_\alpha\cdot K_e}J^\mu_{\pi}(\vec{\underline{k}}_\alpha^\prime,\vec{\underline{k}}_\alpha)\,.
\end{eqnarray}
This prescription resembles the one used already previously~\cite{Biernat:2009my}, which differs in using the electron current instead of $K_e$. In the $k \rightarrow
\infty$ limit both lead to the same result~(\ref{eq:pionformfactor}).
\section{$\rho$ meson}\label{sec:rho}
Next we discuss the treatment of $j_\alpha = 1$ $q\bar q$ bound states within our framework. We will concentrate on the $\rho$ meson, although the formalism is general and applicable to any relativistic spin-1 bound system of 2 equal-mass constituents, with the most prominent example being the deuteron with an instantaneous NN-interaction~\cite{Biernat:2011mp}. The positively charged $\rho$-meson, $\rho^{+}$, is like the $\pi^{+}$, considered within a constituent quark model to be a confined pair of a $u$- and a $\bar{d}$-quark. Before discussing the $\rho$-meson current we note that the Clebsch-Gordan coefficients in the wave functions of Eq.~(\ref{eq:mesonwf}), $C_{\frac12\tilde\mu_q\frac12\tilde\mu_q}^{1\mu_s}$ can be expressed in $(2\times2)$-matrix form as~\cite{Buck:1979ff}
  \begin{eqnarray}\label{eq:CG}
   C_{\frac12\tilde\mu_q\frac12\tilde\mu_{\bar q}}^{1\mu_s}&=&\epsilon_{ \mu_s}^\mu(\vec 0)\left(\sigma_\mu\sfrac{\mathrm i \sigma_2}{\sqrt 2}\right)_{\tilde\mu_q\tilde\mu_{\bar q}}\,,
\label{eq:CGprimeddag}\end{eqnarray}
where $\sigma_\mu=(\mathbf 1,\sigma_1,\sigma_2,\sigma_3)$ with $\sigma_i$ the usual Pauli matrices and $\epsilon_{\mu_s}^\mu(\vec 0)$ the polarization vectors for massive spin-1 particles at rest
\begin{eqnarray}\label{eq:polarvecp}
\epsilon_{1}(\vec0)&\equiv&\epsilon_{1} =-\frac{1}{\sqrt{2}}(0,1,\mathrm i,0)\,,\\\label{eq:polarvecm}
\epsilon_{-1}(\vec 0)&\equiv&\epsilon_{-1}=\frac{1}{\sqrt{2}}(0,1,-\mathrm i,0)\,,\\
\epsilon_{0}(\vec 0)&\equiv&\epsilon_{0} =(0,0,0,1)\,.\label{eq:polarvec0}
                                           \end{eqnarray}
Assuming the $\rho$ meson to be a pure $s$-wave ($l=0$, $\underline{\mu}_\alpha=\mu_s$), using Eq.~(\ref{eq:CG}) for the Clebsch-Gordan coefficients in Eq.~(\ref{eq:Jalpha}) and exploiting Lorentz invariance of the four-vector product $\sigma_\mu\epsilon_{\underline{\mu}_\alpha}^\mu $ we get for the $\rho$-meson current
\begin{widetext}
 \begin{eqnarray} \label{eq:bscurrentVector}
&&J^\mu_{\rho}(\vec{\underline{k}}_{\alpha}',\underline{\mu}_\alpha';\vec{\underline{k}}_{\alpha},\underline{\mu}_\alpha) \nonumber\\&&\quad=\frac{\sqrt{\omega_{\underline k_\alpha}\omega_{\underline k_\alpha^{\prime}}}}{8\pi}
\epsilon^{\ast\sigma}_{\underline{\mu}_\alpha'}(\vec{\underline{k}}_{\alpha}')\epsilon^{\tau}_{\underline{\mu}_\alpha}(\vec{\underline{k}}_{\alpha})[B_c^{-1}(\underline{v}_{\alpha}')]_{\,\,\sigma}^{\lambda}[B_c^{-1}(\underline{v}_{\alpha})]_{\,\,\tau}^{\nu} \int \frac{\mathrm d^3 \tilde k_{q}'}{ \omega_{k_q}}
\sqrt{\frac{m_{q\bar q}}{m_{q\bar q}'}}\sqrt{\frac{\omega_{k_q} +\omega_{k_{\bar q}}}{\omega_{k_q'}+\omega_{k_{\bar q}'}}}
u^\ast_{n0}(|\vec{\tilde{k}}_q'|)u_{n0}(|\vec{\tilde{k}}_q|)
\nonumber\\&&\quad\times
\sum_{ \mu_q, \tilde\mu_q, \dots}(Q_q+Q_{\bar q}) j_q^\mu(\vec{k}_q^\prime,\mu_q^\prime;\vec{k}_q,\mu_q)
 \nonumber\\&&\quad\times D^{\frac12}_{\mu_q\tilde \mu_q}\left[R_W\left(\sfrac {\tilde k_q}{m_q}, B_{c}(v_{q\bar q})\right)\right](\sigma_\nu)_{\tilde \mu_q\tilde \mu_{\bar q}}
    D^{\frac12}_{\tilde{\mu}_{\bar q}\tilde{\mu}'_{\bar q}}\left[R_W\left(\sfrac{\tilde k_{\bar q}'}{m_q}, B_{c}^{-1}(v_{q\bar q}) B_{c} (v'_{q\bar q})\right)\right]
 (\sigma_\lambda)_{\tilde \mu_{\bar q}'\tilde \mu_q'} D^{\frac12}_{\tilde \mu_q'\mu_q'}\left[ R_W^{-1}\left(\sfrac{\tilde k_q'}{m_q}, B_{c}(v_{q\bar q}')\right)\right] \,,\nonumber\\
\end{eqnarray}
where $\underline{v}_{\alpha}^{(\prime)}=\underline{k}_\alpha^{(\prime)}/m_\rho$ is the four-velocity of the confined $q\bar q$ pair in the electron-meson rest frame and $\epsilon^{\mu}_{\underline{\mu}_\alpha}(\vec{\underline{k}}_{\alpha}^{(\prime)})= [B_c(\underline{v}_{\alpha}^{(\prime)})]_{\,\,\nu}^{\mu}\epsilon^{\nu}_{\underline{\mu}_\alpha}(\vec 0)$ is the boosted polarization vector. In the derivation of Eq.~(\ref{eq:bscurrentVector}) we have used the properties of the Wigner $D$-functions together with $\sigma_2 D^{\frac12\ast}(R_W)\sigma_2=D^{\frac12}(R_W)$. It has been shown in Ref.~\cite{Biernat:2011mp} that this current satisfies hermiticity, as in the pion case, i.e.
$[J^\mu_{\rho}(\vec{\underline{k}}_{\alpha}',\underline{\mu}_\alpha';\vec{\underline{k}}_{\alpha},\underline{\mu}_\alpha)]^\dag=J^\mu_{\rho}( \vec{\underline{k}}_{\alpha},\underline{\mu}_\alpha;\vec{\underline{k}}_{\alpha}',\underline{\mu}_\alpha')$.
Furthermore, the current with the correct covariance properties is obtained by going to the physical particle momenta and corresponding spins by means of a canonical-spin boost with overall velocity $\underline v$:
\begin{eqnarray}
 &&J_{\rho}^\mu( \vec p_\alpha^\prime, \sigma_\alpha^\prime; \vec p_\alpha,\sigma_\alpha)\nonumber\\&&\quad=\sum_{\underline{\mu}_\alpha '\underline{\mu}_\alpha}[B_{c}(\underline v)]_{\,\,\nu}^\mu
J_{\rho}^\nu( \vec{\underline{k}}_\alpha^\prime, \underline{\mu}_\alpha^\prime; \vec{\underline{k}}_\alpha,\underline{\mu}_\alpha) D^{1 \ast}_{\underline{\mu}_\alpha '\sigma_\alpha' }[R_W^{-1}(\underline{v}_{\alpha}',B_{ c}(\underline v))]
D^{1}_{\underline{\mu}_\alpha \sigma_\alpha }[R_W^{-1}(\underline{v}_{\alpha},B_{c}(\underline v))]
\nonumber\\&&\quad=\frac{\sqrt{\omega_{\underline{k}_\alpha}\omega_{\underline{k}_\alpha^{\prime}}}}{8\pi}[B_c^{-1}(\underline v)]_{\,\,\kappa}^{\sigma}
\epsilon^{\ast\kappa}_{\sigma_\alpha'}(\vec p_{\alpha}')[B_c^{-1}(\underline v)]_{\,\,\omega}^{\tau}\epsilon^{\omega}_{\sigma_\alpha}(\vec p_{\alpha})[B_c^{-1}(\underline{v}_{\alpha}')]_{\,\,\sigma}^{\lambda}[B_c^{-1}(\underline{v}_{\alpha})]_{\,\,\tau}^{\nu} \int \frac{\mathrm d^3 \tilde k_{q}'}{ \omega_{k_q}}
\sqrt{\frac{m_{q\bar q}}{m_{q\bar q}'}}\sqrt{\frac{\omega_{k_q} +\omega_{k_{\bar q}}}{\omega_{k_q'}+\omega_{k_{\bar q}'}}}
u^\ast_{n0}(|\vec{\tilde{k}}_q'|)u_{n0}(|\vec{\tilde{k}}_q|)
\nonumber\\&&\qquad\times
\sum_{ \mu_q, \tilde\mu_q, \dots}
D^{\frac12}_{\sigma_q\tilde \mu_q}\left[R_W\left(\sfrac {\tilde k_q}{m_q}, B_{c}(\underline v)B_{c}(v_{q\bar q})\right)\right](\sigma_\nu)_{\tilde \mu_q\tilde \mu_{\bar q}}
    D^{\frac12}_{\tilde{\mu}_{\bar q}\tilde{\mu}'_{\bar q}}\left[R_W\left(\sfrac{\tilde k_{\bar q}'}{m_q}, B_{c}^{-1}(v_{q\bar q}) B_{c} (v'_{q\bar q})\right)\right]
 (\sigma_\lambda)_{\tilde \mu_{\bar q}'\tilde \mu_q'}
 \nonumber\\&&\qquad\times D^{\frac12}_{\tilde \mu_q'\sigma_q'}\left[ R_W\left(\sfrac{p_q'}{m_q}, B_{c}^{-1}(v_{q\bar q}')B_{c}^{-1}(\underline v)\right)\right] (Q_q+Q_{\bar q}) j_q^\mu(\vec{p}_q^\prime,\sigma_q^\prime;\vec{p}_q,\sigma_q)\,,\label{eq:bscurrentVectorphys}
\end{eqnarray}

where we have used the transformation properties of the quark current and the $\rho$-meson polarization vectors,
\begin{eqnarray}
 [B_{c}(\underline v)]_{\,\,\nu}^\mu j_q^\mu(\vec{k}_q^\prime,\mu_q^\prime;\vec{k}_q,\mu_q)=\sum_ {\sigma_q'\sigma_q} j_q^\mu(\vec{p}_q^\prime,\sigma_q^\prime;\vec{p}_q,\sigma_q)D^{\frac12\ast}_{\sigma_q'\mu_q'}\left[ R_W^{-1}\left(\sfrac{p_q'}{m_q}, B_{c}^{-1}(\underline v)\right)\right]D^{\frac12}_{\sigma_q \mu_q}\left[R_W^{-1}\left(\sfrac {p_q}{m_q}, B_{c}^{-1}(\underline v)\right)\right]\,,\nonumber\\
 \label{eq:polphysunphy}\end{eqnarray}
 and
 \begin{eqnarray}
\sum_{\underline{\mu}_\alpha}\epsilon^\mu_{\underline{\mu}_\alpha}(\vec{\underline{k}}_\alpha)D_{\underline{\mu}_\alpha\sigma_\alpha}^{1}\left[ R_W^{-1}( \underline{v}_\alpha,B_ c(\underline v))\right]= [B_c^{-1}(\underline v)]^\mu_{\,\,\nu}\epsilon^\nu_{\sigma_\alpha}(\vec p_\alpha)\,,\nonumber\\
\end{eqnarray}
respectively.
\end{widetext}
Current conservation does, in general, not hold for the electromagnetic vector-meson current as given by Eq.~(\ref{eq:bscurrentVectorphys}), i.e.
\begin{eqnarray}\label{eq:currentnonconserv}
(p_\alpha'-p_\alpha)_{\mu}J^\mu_{\rho}(\vec p_{\alpha}',\sigma_\alpha';\vec p_{\alpha},\sigma_\alpha)\neq0\,.
\end{eqnarray}
The formal reason for this failure is the fact that the product of the three Wigner $D$-functions together with the two Clebsch-Gordan coefficients in Eq.~(\ref{eq:bscurrentVectorphys}) \textit{cannot} be written as one single Wigner $D$-function that depends only on consecutive Wigner rotations, as in the pseudoscalar case. Consequently, the properties of the Wigner $D$-functions necessary for showing current conservation cannot be used (for a detailed analysis, which is quite intricate, we refer to App. D.3 of Ref.~\cite{Biernat:2011mp}). The non-vanishing of the left-hand side of Eq.~(\ref{eq:currentnonconserv}) is supported by the analysis  of the covariant structure of the current in the next section and by the numerical results.

Before discussing the current structure, we note that the polarization vectors can be pulled out of the current~(\ref{eq:bscurrentVectorphys}),
\begin{eqnarray}
 J_{\rho}^\mu( \vec p_\alpha^\prime, \sigma_\alpha^\prime; \vec p_\alpha,\sigma_\alpha)=[J_{\rho}( \vec p_\alpha^\prime, \vec p_\alpha)]_{\kappa\omega}^\mu
\epsilon^{\ast\kappa}_{\sigma_\alpha'}(\vec p_{\alpha}')\epsilon^{\omega}_{\sigma_\alpha}(\vec p_{\alpha})\,,\nonumber\\
\end{eqnarray}
leaving a rank-3 Lorentz tensor $[J_{\rho}]_{\sigma\tau}^\mu$ that is independent of the incoming and outgoing spin orientations, $\sigma_\alpha$ and $\sigma_\alpha^\prime$, respectively. This current tensor will be used, alternatively to the current, to extract the form factors in a convenient way. Finally, the microscopic expression of the current tensor can be easily read off from Eq.~(\ref{eq:bscurrentVectorphys}) as
\begin{widetext}
\begin{eqnarray}
 &&[J_{\rho}( \vec p_\alpha^\prime, \vec p_\alpha)]_{\kappa\omega}^\mu\nonumber\\&&\quad=\frac{\sqrt{\omega_{\underline{k}_\alpha}\omega_{\underline{k}_\alpha^{\prime}}}}{8\pi}[B_c^{-1}(\underline v)]_{\,\,\kappa}^{\sigma}
[B_c^{-1}(\underline v)]_{\,\,\omega}^{\tau}[B_c^{-1}(\underline{v}_{\alpha}')]_{\,\,\sigma}^{\lambda}[B_c^{-1}(\underline{v}_{\alpha})]_{\,\,\tau}^{\nu} \int \frac{\mathrm d^3 \tilde k_{q}'}{ \omega_{k_q}}
\sqrt{\frac{m_{q\bar q}}{m_{q\bar q}'}}\sqrt{\frac{\omega_{k_q} +\omega_{k_{\bar q}}}{\omega_{k_q'}+\omega_{k_{\bar q}'}}}
u^\ast_{n0}(|\vec{\tilde{k}}_q'|)u_{n0}(|\vec{\tilde{k}}_q|)
\nonumber\\&&\qquad\times
\sum_{ \mu_q, \tilde\mu_q, \dots}
D^{\frac12}_{\sigma_q\tilde \mu_q}\left[R_W\left(\sfrac {\tilde k_q}{m_q}, B_{c}(\underline v)B_{c}(v_{q\bar q})\right)\right](\sigma_\nu)_{\tilde \mu_q\tilde \mu_{\bar q}}
    D^{\frac12}_{\tilde{\mu}_{\bar q}\tilde{\mu}'_{\bar q}}\left[R_W\left(\sfrac{\tilde k_{\bar q}'}{m_q}, B_{c}^{-1}(v_{q\bar q}) B_{c} (v'_{q\bar q})\right)\right]
 (\sigma_\lambda)_{\tilde \mu_{\bar q}'\tilde \mu_q'}
 \nonumber\\&&\qquad\times D^{\frac12}_{\tilde \mu_q'\sigma_q'}\left[ R_W\left(\sfrac{p_q'}{m_q}, B_{c}^{-1}(v_{q\bar q}')B_{c}^{-1}(\underline v)\right)\right] (Q_q+Q_{\bar q}) j_q^\mu(\vec{p}_q^\prime,\sigma_q^\prime;\vec{p}_q,\sigma_q)\,.\label{eq:bscurrentVectorphys2}
\end{eqnarray}
\end{widetext}
\subsection{Covariant structure of the current}
\label{sec:vectorboundstatesystem}
The correct electromagnetic $\rho$-meson current, which we denote by $I^\mu_{\rho}(\vec p_\alpha^\prime,\sigma_\alpha'; \vec p_\alpha,\sigma_\alpha)$, depends on 3 form factors, for which we choose $F_1$, $F_2$ and $G_M$. These are functions of Mandelstam $t=-Q^2$ only. Its covariant structure is obtained by constructing from the tensor $\epsilon^{\mu\ast}_{\sigma_\alpha'}(\vec p_\alpha')\epsilon^{\nu}_{\sigma_\alpha}(\vec p_\alpha)$ all hermitian, current-conserving four-vectors by appropriate multiplication and contraction with $\mathrm g^{\mu\nu}$, the sum $P_\alpha^\mu$ and/or the difference $d^\mu:=p_\alpha'^\mu-p_\alpha^\mu$ of the incoming and outgoing bound-state four-momenta. However,  as in the pion case, the covariant structure of our $\rho$-meson current (\ref{eq:bscurrentVectorphys}) cannot be solely built from the incoming
and outgoing $\rho$-meson momenta and spins. Due to the violation of cluster separability in the Bakamjian-Thomas framework, we expect that it exhibits an additional dependence on
the sum of the electron momenta $P_e$. Furthermore, unlike the pion case we cannot demand current conservation (cf. Eq.~(\ref{eq:currentnonconserv})) and therefore we have to  allow, in addition, for non-conserved Lorentz structures proportional to $d^\mu$.

The explicit construction of the covariant structure of $J_{\rho}^\mu( \vec p_\alpha^\prime, \sigma_\alpha^\prime; \vec p_\alpha,\sigma_\alpha)$ is discussed in detail in App. D.4 of Ref.~\cite{Biernat:2011mp}. The analysis reveals that one can find 11 hermitian covariants by contracting and/or multiplying the tensor $\epsilon^{\mu\ast}_{\sigma_\alpha'}(\vec p_\alpha')\epsilon^{\nu}_{\sigma_\alpha}(\vec p_\alpha)$ with $\mathrm g^{\mu\nu}$ and/or the available four-vectors $P_\alpha^\mu$, $d^\mu$ and/or $P_e^\mu$. Consequently, we can parametrize $J^\mu_{\rho}$ in terms of 11 form factors, the 3 physical form factors denoted by $f_1$, $f_2$ and $g_M$ and 8 spurious form factors denoted by $b_1,\ldots, b_8$. The form factors exhibit, due to the non-locality of the electromagnetic vertex in the Bakamjian-Thomas framework~\cite{Biernat:2009my}, an additional dependence on Mandelstam $s$ which is expressed, for convenience, in terms of $k$ defined through Eqs.~(\ref{eq:Mandelstams})-(\ref{eq:kmagnitude2}).
Introducing the short-hand notation $\epsilon^{\mu\ast}
_{\sigma_\alpha'}(\vec p_\alpha')\equiv\epsilon^{\prime\mu\ast}$ and $\epsilon_{\sigma_{\alpha}}^\mu (\vec p_\alpha)\equiv \epsilon^\mu$ and dropping the arguments of the form factors which depend on $Q^2$ and $k$, the expansion of the current in terms of the hermitian covariants times the form factors reads

\begin{widetext}

\begin{eqnarray}
&&J_{\rho}^\mu( \vec p_\alpha^\prime, \sigma_\alpha^\prime; \vec p_\alpha,\sigma_\alpha)\nonumber\\&&\quad=
\left[ f_1\epsilon^{\prime\ast}\cdot\epsilon+f_2\frac{(\epsilon^{\prime\ast}\cdot d)
(\epsilon\cdot d)}{2m_{\rho}^2 }\right] P_\alpha^\mu+
g_{M}\left[\epsilon^{\prime\mu\ast}(\epsilon\cdot d)-
\epsilon^\mu(\epsilon^{\prime\ast}\cdot d)\right]\nonumber\\&&\qquad+
\frac{ m_{\rho}^2}{P_e\cdot P_\alpha}\left[
b_1 \epsilon^{\prime\ast}\cdot\epsilon+b_2\frac{(\epsilon^{\prime\ast}\cdot d)
(\epsilon\cdot d)}{m_{\rho}^2 }+
b_3\,4 m_{\rho}^2 \frac{(\epsilon^{\prime\ast}\cdot P_e)(\epsilon\cdot P_e)}{(P_e\cdot P_\alpha)^2}
+ b_4 \frac{ (\epsilon^{\prime\ast}\cdot d)(\epsilon\cdot P_e)-
(\epsilon^{\prime\ast}\cdot P_e)(\epsilon\cdot d) }{P_e\cdot P_\alpha}
\right] P_e^\mu\nonumber\\&&\qquad
+ b_5\,4m_{\rho}^2 \frac{(\epsilon^{\prime\ast}\cdot P_e)(\epsilon\cdot P_e)}{(P_e\cdot P_\alpha)^2}P_\alpha^\mu
+
b_6 \frac{(\epsilon^{\prime\ast}\cdot d)(\epsilon\cdot P_e)-
(\epsilon^{\prime\ast}\cdot P_e)(\epsilon\cdot d) }{P_e\cdot P_\alpha} P_\alpha^\mu
+
b_7 \,2 m_{\rho}^2\frac{\epsilon^{\prime\mu\ast}(\epsilon\cdot P_e)+
\epsilon^\mu(\epsilon^{\prime\ast}\cdot P_e)}{P_e\cdot P_\alpha}
\nonumber\\&&\qquad+b_8
\frac{(\epsilon^{\prime\ast}\cdot d)(\epsilon\cdot P_e)+
(\epsilon^{\prime\ast}\cdot P_e)(\epsilon\cdot d) }{P_e\cdot P_\alpha}\,d^\mu\,.\nonumber\\
\label{eq:covstructurephysdeutcurr}
\end{eqnarray}
As in the pion case there seems to be an ambiguity how to separate the physical from unphysical contributions. Again it turns out, however, that only the above decomposition with the associated definition of form factors gives the correct charge of the bound state at zero momentum transfer, as it is required for the physical charge form factor $G_C$. Again this justifies to define unphysical (or spurious) contributions as structures proportional to first or higher powers of $P_e$. These are the structures multiplied by the spurious form factors $b_1,\,\ldots,\,b_8$ in Eq.~(\ref{eq:covstructurephysdeutcurr}).
By separation of the polarization vectors we find the covariant structure of the current tensor:
 \begin{eqnarray}
\label{eq:currentphneospin1cov}
&&[J_{\rho}(\vec p_{\alpha}',\vec p_{\alpha})]^\mu_{\sigma\tau}\nonumber\\&&\quad=
\left( f_1\mathrm g_{\sigma\tau}+f_2\frac{d_\sigma d_\tau}{2m_{\rho}^2 }\right) P_\alpha^\mu
+g_{M}\left(\mathrm g^\mu_{\sigma}d_\tau-\mathrm g^\mu_{\tau}d_\sigma\right)+
\left[
b_1 \mathrm g_{\sigma\tau}+b_2\frac{d_\sigma d_\tau}{m_{\rho}^2 }
+b_34 m_{\rho}^2 \frac{P_{e \sigma}P_{e \tau}}{(P_e\cdot P_\alpha)^2}
+b_4 \frac{ P_{e \tau}d_\sigma-  P_{e \sigma}d_\tau}{P_e\cdot P_\alpha}
\right]\frac{ m_{\rho}^2}{P_e\cdot P_\alpha} P_e^\mu\nonumber\\&&\qquad+
b_5 \frac{P_{e \sigma}P_{e \tau}}{(P_e\cdot P_\alpha)^2}\,4 m_{\rho}^2P_\alpha^\mu
+b_6 \frac{ P_{e \tau}d_\sigma-  P_{e \sigma}d_\tau}{ P_e\cdot P_\alpha} P_\alpha^\mu
+
b_7 \,2 m_{\rho}^2\frac{\mathrm g^\mu_\sigma P_{e\tau}+\mathrm g^\mu_\tau P_{e\sigma}}{P_e\cdot P_\alpha}
+b_8
\frac{d_\sigma P_{e\tau}+d_\tau P_{e\sigma} }{P_e\cdot P_\alpha}\,d^\mu\,.
\end{eqnarray}
The covariant structures of $J_{\rho}^\mu( \vec p_\alpha^\prime, \sigma_\alpha^\prime; \vec p_\alpha,\sigma_\alpha)$ and $[J_{\rho}(\vec p_{\alpha}',\vec p_{\alpha})]^\mu_{\sigma\tau}$ resemble the corresponding ones obtained within the covariant
light-front approach of Refs.~\cite{Karmanov:1994ck,Carbonell:1998rj} and thus we have adopted their notation and normalizations for the form factors. In this work the authors encounter 8 spurious contributions to the current that are associated with $\omega^\mu$. These $\omega$-dependent spurious contributions correspond to our $K_e$-dependent spurious contributions. As in the pseudoscalar case, the spurious contributions of our vector-meson current can be traced back to the violation of
cluster separability. The $\omega$-dependent contributions of
Ref.~\cite{Carbonell:1998rj} are rather the consequence of
the most general ansatz for their current which has to include the orientation of the light front.
\end{widetext}
\subsection{ Electromagnetic form factors}
\subsubsection{$F_1$,  $F_2$ and $G_M$}
The numerical analysis -- the dynamical ingredients of which will be discussed in the next section --  with the standard kinematics as introduced in Eq.~(\ref{eq:momentumscatt}) ($\vec {\underline v}=0$ and thus $p_\alpha^{(\prime)}=\underline{k}_\alpha^{(\prime)}$ and $\sigma_\alpha^{(\prime)}=\underline{\mu}_\alpha^{(\prime)}$) reveals that the microscopic expression $J^\mu_{\rho}$ of Eq.~(\ref{eq:bscurrentVector}) has indeed 11 independent, non-vanishing matrix elements. Using the short-hand notation $J_{\rho}^\mu( \vec{\underline{k}}_\alpha^\prime, \underline{\mu}_\alpha^\prime; \vec{\underline{k}}_\alpha,\underline{\mu}_\alpha)\equiv J^\mu_{\underline{\mu}_\alpha'\underline{\mu}_\alpha}$ they are given by $J^0_{1-1},\, J^3_{1-1},\,J^0_{00},\, J^3_{00},\, J^0_{10}, \,J^1_{10},
 \,J^2_{10}, \,J^3_{10}, \,J^0_{11},\, J^2_{11}$ and $J^3_{11}$, with all the remaining non-vanishing matrix elements related to them due to parity and time-reversal invariance. Out of the 11 form factors only the 3 physical form factors $f_1$, $f_2$ and $g_M$ are of interest. We extract them from Eq.~(\ref{eq:bscurrentVector}) by using the decomposition~(\ref{eq:covstructurephysdeutcurr}). As in the  pion case we take the limit $k \rightarrow\infty$, where the form factors become independent of $k$, and denote the limiting expressions by capital letters
\begin{eqnarray}
 F_i(Q^2)&=&\lim_{k\rightarrow\infty}f_i(Q^2,k) \quad \text{with} \quad i=1,2\,,\\
 G_M(Q^2)&=&\lim_{k\rightarrow\infty}g_M(Q^2,k)
 \end{eqnarray}
 and
 \begin{eqnarray}
 B_j(Q^2)=\lim_{k\rightarrow\infty}b_j(Q^2,k) \quad \text{with} \quad j=1,\ldots,8\,.
\end{eqnarray}
 Furthermore, using our standard kinematics, we observe that the zeroth and third components of the current become identical in this limit, i.e. $J^0_{\underline{\mu}_\alpha'\underline{\mu}_\alpha}\!\!\stackrel{ k \rightarrow\infty}{\longrightarrow}\!\! J^3_{\underline{\mu}_\alpha'\underline{\mu}_\alpha}$,
 which reduces the number of independent matrix elements from 11 to 7. This means, however, that 4 of the 8 spurious contributions cannot be eliminated by simply taking the limit $k \rightarrow
\infty$. These are the ones connected with $B_5,\ldots,B_8$. As in the pion case, taking the limit $k \rightarrow\infty$ in our formalism resembles the situation in usual light-front dynamics with $\omega=(1,0,0,-1)$, in which the extraction of the form factors is based on the plus component of the current operator. By restricting to the plus component in the usual approach the 4 spurious contributions containing $B_1,\ldots,B_4$, which are proportional to $\omega^\mu$, are eliminated. However, the contributions associated with $B_5,\ldots,B_8$ survive, similar as in our case. In particular, using our standard kinematics a numerical analysis shows that in the limit $k \rightarrow\infty$ there is a non-vanishing contribution to the $\mu=1$ component of the current, namely
\begin{eqnarray}
 \lim_{k\rightarrow\infty} J^1_{10}=-B_7(Q^2)\frac {m_\rho}{\sqrt {2}}+B_8(Q^2) \frac{Q^2}{2\sqrt{2}m_{\rho}}\,,
\end{eqnarray}
which clearly violates current conservation (in our standard kinematics $q^\mu=(0,Q,0,0)^\mu$). This complication occurs due to the increased complexity of spin-1 bound systems as compared to the spin-0 case.

The physical $\rho$-meson current, denoted by $I^\mu_{\sigma_\alpha'\sigma_\alpha}$, that satisfies all required properties should depend only on the 3 physical form factors $F_1$, $F_2$ and $G_M$. Hence the 4 matrix elements $I^0_{11}, I^0_{1-1}, I^0_{10}$ and $I^0_{00}$, which cannot be related by parity or time-reversal invariance, should satisfy the, so-called, \textit{angular condition} (see, e.g., Refs.~\cite{Grach:1983hd,Carbonell:1998rj,Bakker:2002aw})
\begin{eqnarray}\label{eq:angularcondition}
(1+2\eta)I^0_{11}+I^0_{1-1}-2\sqrt{2\eta}I^0_{10}-I^0_{00}=0\,,
\end{eqnarray}
where $\eta=\frac{Q^2}{4 m_{\rho}^2}$. For our current matrix elements $J^0_{\sigma_\alpha'\sigma_\alpha}$, the angular condition is, however, not satisfied -- not even in the limit $k \rightarrow
\infty$ -- due to the spurious contributions, in particular the non-vanishing form factors $B_5$ and $B_7$:
  \begin{eqnarray}\label{eq:angularconditionviolation}&&
\lim_{k\rightarrow \infty}\frac{1}{2k}\left[(1+2\eta)J^0_{11}+J^0_{1-1}-2\sqrt{2\eta}J^0_{10}-J^0_{00}\right]
\nonumber\\&&\;\;\;\;=-\left[B_5(Q^2) +B_7(Q^2)\right]\,.
\end{eqnarray}
It is thus not possible to extract the 3 physical form factors in an unambiguous way from $J^0_{11},J^0_{1-1},J^0_{10}$ and $J^0_{00}$, unless the spurious parts of these current matrix elements are first separated. This problem occurs in the usual light-front approach as well (with the $0$ replaced by the $+$ component of the current). Different triplets of current matrix elements have been chosen in the literature to calculate the 3 physical form factors~\cite{Grach:1983hd,Chung:1988my,Brodsky:1992px,Frankfurt:1993ut}. These different prescriptions lead, in general, to different results for the form factors as soon as the angular condition is violated, an example being the naive impulse approximation. For a numerical and analytical comparison of different approaches see  Refs.~\cite{Cardarelli:1994yq,Carbonell:1998rj,Karmanov:1996qc}, respectively.

If the angular condition were satisfied, the different prescriptions would lead to the same form factor results~\cite{Keister:1993mg,Carbonell:1998rj}. However, the magnetic form factor obtained from the plus component of the current could still contain the spurious form factor $B_6$ depending on which current matrix elements are used to calculate $G_M$. Translating to our case this means, e.g., that
\begin{eqnarray}
\lim_{k\rightarrow\infty}\frac1k\left(J^0_{10}-\frac{m_\rho\sqrt{2}}{Q}J^0_{11}\right)&=& G_M(Q^2)-B_6(Q^2)\,,\label{eq:GMB6}
\end{eqnarray}
which implies that $G_M$ \emph{cannot} be directly extracted from the matrix elements $J^0_{10}$ and $J^0_{11}$.  As we shall see later by a numerical analysis, the spurious contributions are altogether relatively small, such that the different prescriptions which do not separate them lead to rather similar results.

Nonetheless, we have found an unambiguous way to cleanly separate the physical from the unphysical contributions: a careful analysis of the current matrix elements reveals that in the limit $k\rightarrow\infty$ the 3 matrix elements $J^0_{11}, J^0_{1-1}$ and $J^2_{11}$ do not contain spurious contributions in the leading order of a $1/k$ expansion  (a similar analysis in the light-front formalism can be found in Ref.~\cite{Melikhov:2001pm}). These \lq\lq good'' matrix elements are therefore appropriate for the extraction of the physical form factors, which are then given by (for a derivation we refer to App. F of Ref.~\cite{Biernat:2011mp})
\begin{widetext}

\begin{eqnarray}
 \label{eq:ff1me}
F_1(Q^2)&=& -\lim_{k\rightarrow\infty} \frac{1}{2k}\left[J^0_{11}+J^0_{1-1}\right]=-\frac{1}{4\pi}\int\mathrm{d}^3\tilde{k}'_q\sqrt{\frac{m_{q\bar q}}{m'_{q\bar q}}}u_{n0}^\ast\left(|\vec{\tilde{k}}_q'|\right)u_{n0}\left(|\vec{\tilde{k}}_q| \right)
(\mathcal S_{11}^{+}+\mathcal S_{1-1}^{+})\,,
\\
\label{eq:ff2me}
F_2(Q^2)&=&-\frac{1}{\eta}\lim_{k\rightarrow\infty} \frac{1}{2k}J^0_{1-1}=-\frac{1}{4\pi\eta}\int\mathrm{d}^3\tilde{k}'_q\sqrt{\frac{m_{q\bar q}}{m'_{q\bar q}}}u_{n0}^\ast\left(|\vec{\tilde{k}}_q'|\right)u_{n0}\left(|\vec{\tilde{k}}_q|\right)\mathcal S_{1-1}^+\,
\end{eqnarray}
and
\begin{eqnarray}
\label{eq:ffGMme}
 G_M(Q^2)&=&-\frac{\mathrm i}{Q} \lim_{k\rightarrow\infty}J^2_{11}=
-\frac{\mathrm i}{4\pi Q}\int\mathrm{d}^3\tilde{k}'_q\sqrt{\frac{m_{q\bar q}}{m'_{q\bar q}}}u_{n0}^\ast\left(|\vec{\tilde{k}}_q'|\right)u_{n0}\left(|\vec{\tilde{k}}_q|\right)
\frac{m_{q\bar q}'}{ ( m_{q\bar q}' + 2  \tilde k_q'^3)}
\left(
\tilde k_q'^2 \mathcal S_{11}^{+}+  \frac{\mathrm i \,Q}{2}
\mathcal S_{11}^{-}\right)\,.
\end{eqnarray}
Here $\mathcal S_{\underline{\mu}_\alpha'\underline{\mu}_\alpha}^{+}$ and $\mathcal S_{\underline{\mu}_\alpha'\underline{\mu}_\alpha}^{-}$ are the spin rotation factors
\begin{eqnarray}\label{eq:S1}
\mathcal S^{\pm}_{\underline{\mu}_\alpha'\underline{\mu}_\alpha}&:=&\lim_{k\rightarrow \infty}\frac12\sum_{ \mu_q, \tilde{\mu}_q,\ldots}(\pm 1)^{\mu_q-\frac12}D^{\frac12}_{\mu_q\tilde{\mu}_q}\left[ R_W\left(\sfrac{\tilde k_q}{m_q},B_c\left(v_{q\bar q} \right)\right)\right]\left(\vec{\epsilon}_{\underline{\mu}_\alpha}\cdot \vec{\sigma}\right)_{\tilde{\mu}_q \tilde{\mu}_{\bar q}}
D^{\frac12}_{\tilde{\mu}_{\bar q}\tilde{\mu}_{\bar q}'}\left[R_W\left(\sfrac{\tilde k_{\bar q}'}{m_q},B_c^{-1}\left(v_{q\bar q}\right)B_c\left(v'_{q\bar q}\right)\right)\right] \nonumber\\&&\times \left(\vec\epsilon^{\ast}_{\underline{\mu}_\alpha'}\cdot \vec \sigma\right)_{\tilde{\mu}_{\bar q}' \tilde{\mu}_{q}'}D^{\frac12}_{\tilde{\mu}_q'\mu_q}\left[R_W\left(\sfrac{k_q'}{m_q},B_c^{-1}\left(v'_{q\bar q} \right)\right)\right]\,\,,
\end{eqnarray}
where $\vec\epsilon_{\underline{\mu}_\alpha}$ and $\vec\epsilon^{\ast}_{\underline{\mu}_\alpha'}$ are the spin-1 polarization three-vectors in the rest frame, Eqs.~(\ref{eq:polarvecp})-(\ref{eq:polarvec0}).

Another, equivalent prescription proposed in the covariant light-front approach of Refs.~\cite{Karmanov:1994ck,Carbonell:1998rj} to extract the physical form factors, can also be applied to our case due to the similarities between both approaches. To this end we define appropriate tensors $F_{1\mu}^{\sigma\tau}$, $F_{2\mu}^{\sigma\tau}$ and $G_{M\mu}^{\sigma\tau}$ that project out the form factors $F_1$, $F_2$ and $G_M$ from the current tensor $[J_\rho (\vec p_{\alpha}',\vec p_{\alpha})]^\mu_{\sigma\tau}$. These projection tensors, fixed by the decomposition~(\ref{eq:currentphneospin1cov}), read~\cite{Karmanov:1994ck,Carbonell:1998rj}:
\begin{eqnarray}
F_{1\mu}^{\sigma\tau}&:=&
 \frac{P_{e\mu}}{ P_e\cdot P_\alpha }\left(\mathrm g^{\sigma\tau}-\frac{d^\sigma d^\tau}{d^2}-\frac{P_\alpha ^\sigma P_e^\tau+P_\alpha ^\tau P_e^\sigma}{P_e\cdot P_\alpha }+P_\alpha ^2
\frac{P_e^\sigma P_e^\tau}{(P_e\cdot P_\alpha )^2}\right)\,,\\
F_{2\mu}^{\sigma\tau}&:=&-\frac{P_{e\mu}}{ (P_e\cdot P_\alpha )d^2}\left(\mathrm g^{\sigma\tau}-\frac{d^\sigma d^\tau}{d^2}-
\frac{P_\alpha ^\sigma P_e^\tau+P_\alpha ^\tau P_e^\sigma}{P_e\cdot P_\alpha }
\right.+4 m_{\alpha }^2\frac{P_e^\sigma P_e^\tau}{ (P_e\cdot P_\alpha )^2}-\left.\frac{d^\sigma P_e^\tau-d^\tau P_e^\sigma}{P_e\cdot P_\alpha }\right)\,
\end{eqnarray}
and
\begin{eqnarray}
 G_{M\mu}^{\sigma\tau}&:=&\frac12
\left[\frac{\mathrm g^\sigma_\mu d^\tau-\mathrm g^\tau_\mu d^\sigma}{d^2}+\frac{\mathrm g^\sigma_\mu P_e^\tau+\mathrm g^\tau_\mu P_e^\sigma}{P_e\cdot P_\alpha }+
P_{\alpha \mu}\left(\frac{d^\sigma P_e^\tau-d^\tau P_e^\sigma}{d^2(P_e\cdot P_\alpha )}-
2\frac{P_e^\sigma P_e^\tau}{(P_e\cdot P_\alpha )^2}\right)
-d_{\mu} \frac{d^\sigma P_e^\tau+d^\tau P_e^\sigma}{(P_e\cdot P_\alpha )d^2}\right.\nonumber\\&&\left.\;\;\;\;\;\;\;\;+\frac{P_{e\mu}}{P_e\cdot P_\alpha }\left(-P_\alpha ^2\frac{d^\sigma P_e^\tau-d^\tau P_e^\sigma}{(P_e\cdot P_\alpha )d^2}+
\frac{d^\sigma P_\alpha ^\tau-d^\tau P_\alpha ^\sigma}{d^2}+
2P_\alpha ^2\frac{P_e^\sigma P_e^\tau}{(P_e\cdot P_\alpha )^2}-
\frac{P_\alpha ^\sigma P_e^\tau-P_\alpha ^\tau P_e^\sigma}{P_e\cdot P_\alpha }\right)\right]\,.
\end{eqnarray}
Then the form factors are obtained by contraction of the projection tensors with the current tensor:
\begin{eqnarray}\label{eq:ff1projection}
 F_i(Q^2)=\lim_{k\rightarrow \infty}F_{i\mu}^{\sigma\tau}  [J_\rho (\vec p_{\alpha }',\vec p_{\alpha })]^\mu_{\sigma\tau} \quad\text{with}\quad i=1,\,2\quad \text{and}\quad
G_M(Q^2)=\lim_{k\rightarrow \infty}G_{M\mu}^{\sigma\tau} [J_\rho (\vec p_{\alpha }',\vec p_{\alpha })]^\mu_{\sigma\tau}\,.\label{eq:ffGMprojection}
\end{eqnarray}
These are finite expressions and independent of $k$. It has been shown that they are identical to the corresponding ones obtained from the current matrix elements in Eqs.~(\ref{eq:ff1me}),~(\ref{eq:ff2me}) and (\ref{eq:ffGMme}). The physical current that has all required properties is then
\begin{eqnarray}
 I^\mu_{\sigma_\alpha'\sigma_\alpha}= \lim_{k\rightarrow \infty} \left\lbrace \left[F_1(Q^2) (\epsilon^{\prime\ast}\cdot\epsilon) +F_2(Q^2) \frac{(\epsilon^{\prime\ast}\cdot d)
(\epsilon\cdot d)}{2m_{\rho}^2 }\right] P_\alpha^\mu+
G_{M}(Q^2) \left[\epsilon^{\prime\mu\ast}(\epsilon\cdot d)-
\epsilon^\mu(\epsilon^{\prime\ast}\cdot d)\right]\right\rbrace\, ,
\end{eqnarray}
where the form factors $F_1$, $F_2$ and $G_{M}$ are given by Eqs.~(\ref{eq:ff1me})-(\ref{eq:ffGMme}) and $\epsilon^{\prime\ast}=\epsilon^{\ast}_{\sigma_\alpha'}(\vec p_\alpha')$ and $\epsilon=\epsilon_{\sigma_\alpha}(\vec p_\alpha)$.
\end{widetext}

Before we proceed we briefly point out the differences between our and other approaches that use the point form of relativistic dynamics for the calculation of elastic form factors of spin-1 two-particle bound states. The point form has been used in Ref.~\cite{Allen:2000ge} to calculate the deuteron elastic form factors. In this work, similar as in the point-form spectator model~\cite{Wagenbrunn:2000es,Boffi:2001zb}, the BT construction is just applied to obtain the bound-state wave function and its mass, but \emph{not} to calculate the bound-state current as in our case. Instead, a general Lorentz-covariant \emph{ansatz} is made for the bound-state current with the wave function and the bound-state mass serving as inputs. The advantage of this procedure is that cluster separability is trivially satisfied~\cite{Keister:1991sb}. Reference~\cite{Allen:2000ge} makes use of a spectator approximation and the fact that, after imposing current conservation, the number of independent non-vanishing current matrix 
elements in the Breit frame is equal to the number of physical form factors which, in this way, are uniquely determined~\cite{Klink:1998qf}.

Due to the kinematic nature of Lorentz transformations in the point form the current can then be transformed into arbitrary frames. This procedure ensures automatically that the angular condition is satisfied.  In all the point-form approaches  the four-momentum $q^\mu$ transferred to the constituent that is struck by the photon is not the same as the four-momentum transfer $\underline{q}^\mu$ between incoming and outgoing bound state. But, due to the different procedures for deriving the current, our $q^\mu$  differs also from the one in the point-form approaches of Refs.~\cite{Allen:2000ge,Wagenbrunn:2000es,Boffi:2001zb}. As a consequence we get different boosts, Wigner rotations and kinematical factors which explains partly why we are closer to light-front results than to point-form calculations along the lines of Refs.~\cite{Allen:2000ge,Wagenbrunn:2000es,
Boffi:2001zb}.

\subsubsection{$G_C$, $G_Q$ and elastic scattering observables}
The charge and quadrupole form factors $G_C$ and $G_Q$, respectively, are expressed through $F_1$, $F_2$ and $G_M$ by
 \begin{eqnarray}
 &&G_C(Q^2)\nonumber\\&&\quad= -F_1(Q^2)-\frac{2\eta}{3}\left[F_1(Q^2)+G_M(Q^2)-F_2(Q^2)(1+\eta)\right]\nonumber\\\label{eq:rhoGC}
\end{eqnarray}
and
 \begin{eqnarray}
G_Q(Q^2)=-F_1(Q^2)-G_M(Q^2)+F_2(Q^2)(1+\eta)\,.\nonumber\\
\label{eq:rhoGQ}
\end{eqnarray}
These form factors have the limits
\begin{eqnarray}
\lim_{Q^2\rightarrow0}G_C(Q^2)&=&1\,,\\\label{eq:GCrho0}
\lim_{Q^2\rightarrow0}G_M(Q^2)&=&\mu_\rho\,,\label{eq:murho}\\
\lim_{Q^2\rightarrow0}G_Q(Q^2)&=&Q_\rho\,,\label{eq:qrho}
\end{eqnarray}
where $+1$ is the charge in units of the fundamental charge $|\,e\,|$, $\mu_\rho$ the magnetic dipole moment in units of $|\,e\,|/2m_{\rho}$ and $Q_\rho$ the electric quadrupole moment in units of $|\,e\,|/m_{\rho}^2$.
For point-like spin-1 systems the magnetic dipole and the electric quadrupole moments are $\mu_{\text{point}}=2$ and $Q_{\text{point}}=-1$, respectively.

For the discussion of the high $Q^2$-behavior it is useful to switch to the usual observables of elastic electron-$\rho$-meson scattering, which are the structure functions $A(Q^2)$, $B(Q^2)$ and the tensor polarization $T_{20}(Q^2)$. $A(Q^2)$, $B(Q^2)$ are determined from the unpolarized laboratory frame differential cross section using the Rosenbluth formula. We have 
\begin{eqnarray}\label{eq:Aobservable}
A(Q^2)=G_C^2(Q^2)+\frac{8}{9}\eta^2G_Q^2(Q^2)+\frac23 \eta\, G_M^2(Q^2)
\end{eqnarray}and

\begin{eqnarray}
B(Q^2)=\frac43 \eta (1+\eta)G_M^2(Q^2)\,.\label{eq:Bobservable}
\end{eqnarray}
The observable $T_{20}(Q^2)$ for quadrupole polarization is extracted from the difference
in the cross sections for target $\rho$-meson having canonical spin polarizations $\sigma_\alpha=1$ and $\sigma_\alpha=0$.
In terms of form factors it reads
\begin{widetext}
\begin{eqnarray}\label{eq:T20}
T_{20}(Q^2)=-\sqrt 2 \eta\,  \frac{\frac49 \eta \, G_Q^2(Q^2)+\frac43 G_Q(Q^2)G_C(Q^2)+\frac13(\frac12+(1+\eta)\tan^2(\theta/2))G_M^2(Q^2)}{A(Q^2)+B(Q^2)\tan^2(\theta/2)}\, ,
\end{eqnarray}
where $\theta$ is the electron scattering angle in the  laboratory frame.

\vspace{1.0cm}

\end{widetext}

\section{Numerical results}\label{sec:numericalresults}
\subsection{$\rho$-meson wave function}\label{subsec:wf}
For the numerical study of the $\rho$-meson form factors we obviously have to specify the $q\bar q$ bound-state wave function, the constituent-quark masses and the $\rho$-meson mass. For the $\rho$-meson wave function we take a simple harmonic-oscillator form:
\begin{eqnarray}
 \label{eq:u00}
u_{00}(|\vec{\tilde{k}}_q|) = \frac{2}{\pi^{\frac14}a^{\frac32}}\exp\left(-\sfrac{\vec{\tilde{k}}_q^2}{2a^2}\right)\,.
\end{eqnarray}
Such a wave function has also been used in light-front calculations~\cite{Choi:2004ww,Jaus:2002sv} which we want to compare with. The numerical values for the oscillator parameter and the constituent-quark masses will thus be taken from these papers. For the $\rho$-meson mass we will adopt its physical value $m_\rho=0.77~\mathrm{GeV}$.

In addition we will also use our own parametrization. The wave function (\ref{eq:u00}) can be considered as the eigenfunction of a mass-eigenvalue problem in which a harmonic-oscillator confinement potential is added to the {\em square} of the free $q\bar{q}$ mass operator~\cite{KrassniggDiss:2001,Krassnigg:2003gh}.
The mass-eigenvalues are then
\begin{eqnarray}\label{eq:spectrumho}
m_{nl}=\sqrt {8a^2\left(2n+l+\frac32\right)+4m_q+c_0}\, ,
 \end{eqnarray}
where we have allowed for an additional parameter $c_0$ to shift the spectrum.
With $m_q=m_u=m_d=0.34$~GeV prefixed and the remaining 2 parameters chosen in such a way that $m_{00}$ and $m_{10}$ agree with the masses of the ground and first excited state of the $\rho$ meson one finds $a=0.312~\mathrm{GeV}$ and $c_0=-1.04~\mathrm{GeV}^2$~\cite{KrassniggDiss:2001,Krassnigg:2003gh}. The lowest eigenvalue  $m_{00}$ agrees then with the $\rho$-meson mass and the first and second radial excitations are about 10\% too high as compared to experiment.
\subsection{Predictions}
\label{subsec:prediction}
\begin{figure}[t!]
\begin{flushright}
\includegraphics[width=8.6cm]{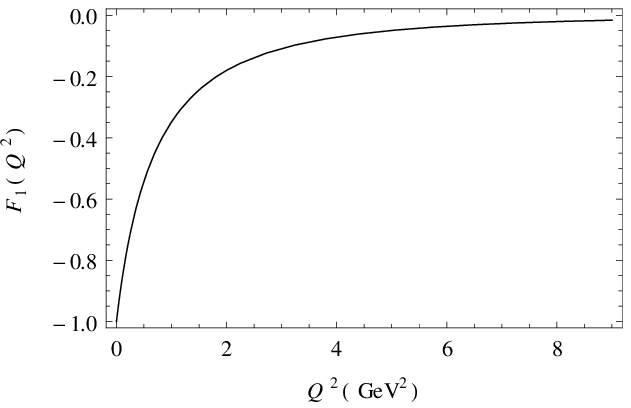}\vspace{.7cm}
\includegraphics[width=8.3cm]{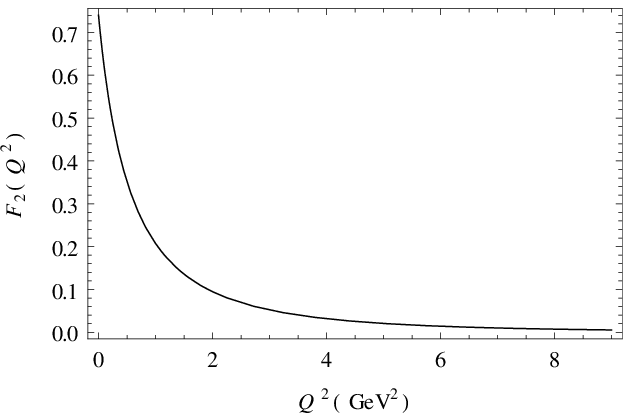}\vspace{.7cm}
\includegraphics[width=8.3cm]{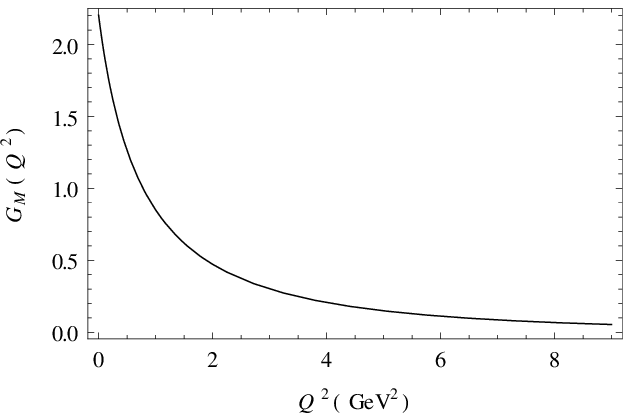}
\caption{The electromagnetic $\rho$-meson form factors
  $F_1\left(Q^2\right)$ (top panel), $F_2\left(Q^2\right)$ (middle panel) and $G_M\left(Q^2\right)$ (bottom panel) for the harmonic-oscillator wave function, Eq.~(\ref{eq:u00}), and parameters $m_q=0.34\, \mathrm {GeV}$, $a=0.312\, \mathrm {GeV}$ and $m_\rho= m_{00}=0.77\, \mathrm {GeV}$.
}\label{fig:rhoFF}                   \end{flushright}
\end{figure}

\begin{figure}
\begin{flushright}
\includegraphics[width=8.3cm]{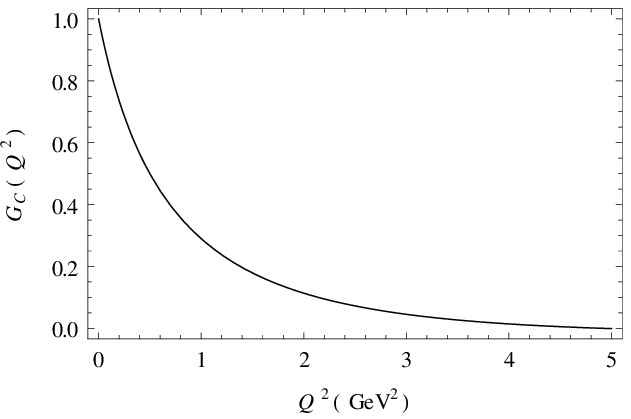}\vspace{.7cm}
\includegraphics[width=8.6cm]{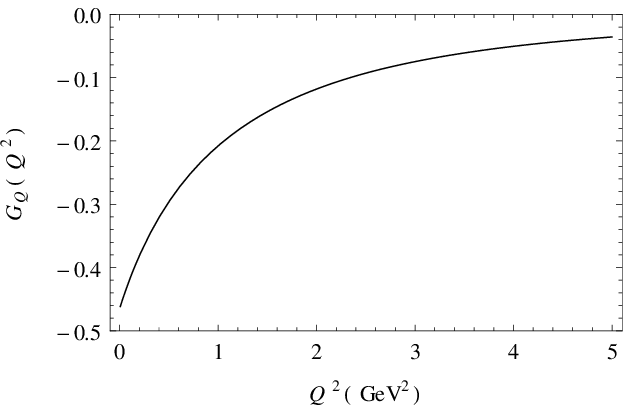}
\caption{The electric charge form factor
  $G_C\left(Q^2\right)$ (top panel) and electric quadrupole form factor
  $G_Q\left(Q^2\right)$ (bottom panel) of the $\rho$-meson evaluated with the same parameters as
in Fig.~\ref{fig:rhoFF}.
}  \label{fig:rhoGCGQ}                    \end{flushright}
\end{figure}

\begin{figure}[t!]
\begin{flushright}

\includegraphics[width=8.3cm]{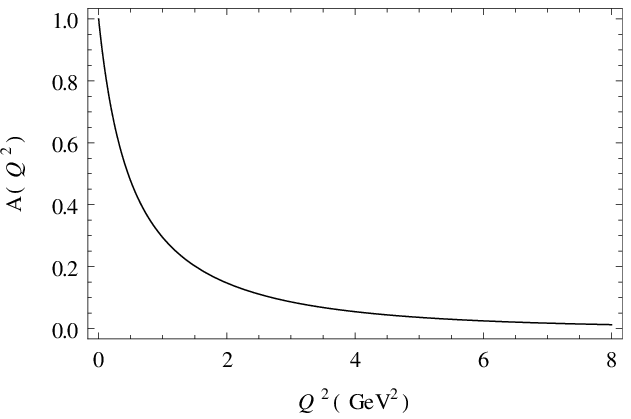}\vspace{.7cm}
\includegraphics[width=8.3cm]{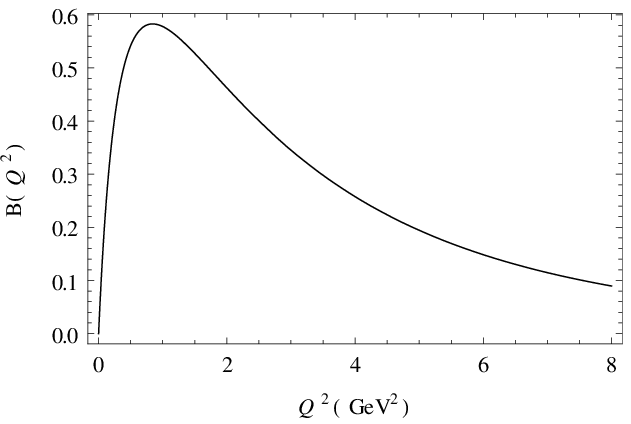}\vspace{.7cm}
\includegraphics[width=8.6cm]{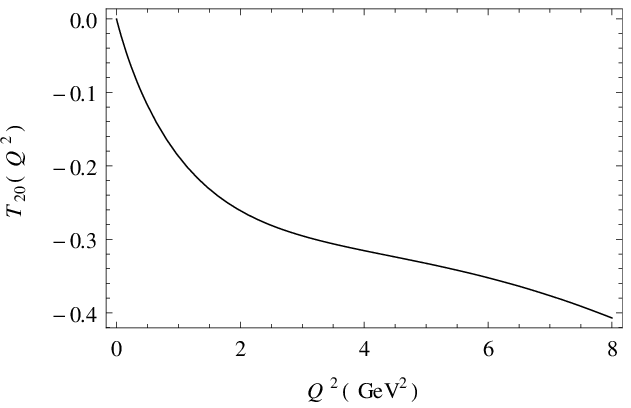}

\caption{The elastic scattering observables $A(Q^2)$ (top panel), $B(Q^2)$ (middle panel) and $T_{20}(Q^2)$ (bottom panel)  evaluated by means of Eqs.~(\ref{eq:Aobservable})--(\ref{eq:T20}) with the same parameters as
in Fig.~\ref{fig:rhoFF}.
}  \label{fig:elastscatt}                    \end{flushright}
\end{figure}

 Numerical results for the electromagnetic $\rho$-meson form factors~(\ref{eq:ff1me})-(\ref{eq:ffGMme}) evaluated with $m_u=m_d=0.34$~GeV, $a=0.312~\mathrm{GeV}$ and ${m_\rho=}m_{00}=0.77~\mathrm{GeV}$ are depicted in Fig.~\ref{fig:rhoFF}.
The corresponding electric charge and quadrupole form factors of the $\rho$ meson, $G_C\left(Q^2\right)$ and $G_Q\left(Q^2\right)$, respectively, are plotted in Fig.~\ref{fig:rhoGCGQ}.
From the top panel in Fig.~\ref{fig:rhoGCGQ} we read off the correct $\rho$-meson charge $G_C(0)=1$ in units of the fundamental charge $|\,e\,|$. This is ensured by the decomposition of the $\rho$-meson current  introduced in Eq.~(\ref{eq:covstructurephysdeutcurr}), which justifies this particular way of separating the physical from the unphysical contributions. Our predictions for the magnetic dipole and the electric quadrupole moment, which are the $Q^2\rightarrow 0$ limits of $G_M\left(Q^2\right)$ and $G_Q\left(Q^2\right)$ (cf. Eqs.~(\ref{eq:murho}) and~(\ref{eq:qrho})), are $\mu_\rho=2.2$ and $Q_\rho=-0.47$ (in units $|\,e\,|/2m_{\rho}$ and $|\,e\,|/m_{\rho}^2$), respectively.
The results for the elastic scattering observables $A(Q^2)$, $B(Q^2)$ and $T_{20}(Q^2)$ are
depicted in Fig.~\ref{fig:elastscatt}.

\begin{figure}[ht!]
\begin{flushright}
\includegraphics[width=8.3cm]{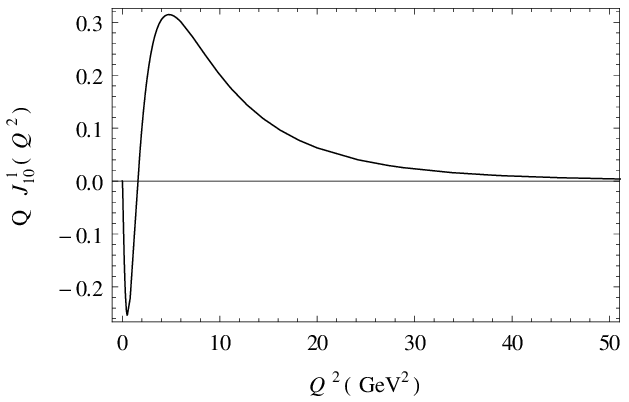}\vspace{.7cm}\\
\includegraphics[width=8.3cm]{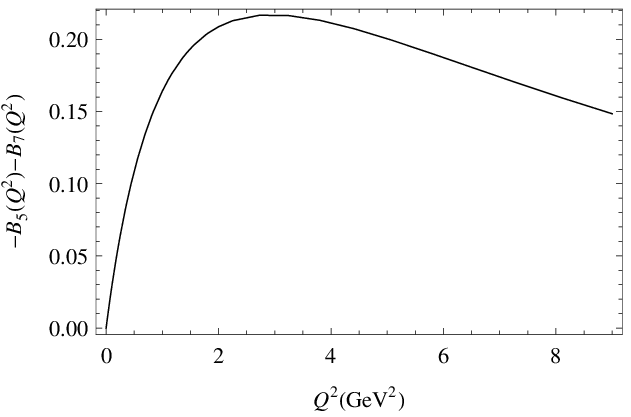}\vspace{.7cm}\\
\includegraphics[width=8.3cm]{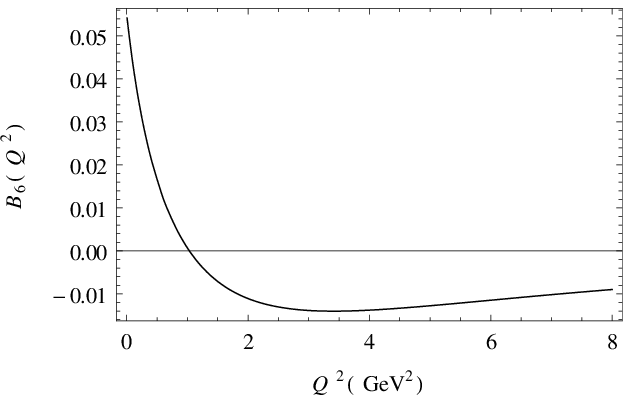}
\caption{\label{fig:spuriousconstrib} The spurious contributions to the $\rho$-meson current. The violation of current conservation given by $Q J^1_{10}$ (top panel), the violation of the angular condition (middle panel) given by Eq.~(\ref{eq:angularconditionviolation}) and the spurious form factor $B_6(Q^2)$ (bottom panel), all calculated with the same parameters as
in Fig.~\ref{fig:rhoFF}.
}
\end{flushright}
\end{figure}

It is also interesting to see how large the spurious contributions to the $\rho$-meson current are that emerge from wrong cluster properties within our approach. Of particular interest are effects violating current conservation and the angular condition as well as the spurious form factor $B_6$ that is relevant in the context of $G_M$. The $\rho$-meson current is not conserved if the 1-component of the current does not vanish when using our standard kinematics of momentum transfer in the 1-direction. A measure for the violation of current conservation is $\lim_{s\rightarrow\infty}q_\mu  J_\rho^\mu \sim Q J^1_{10} (Q^2)$. This quantity is depicted in the top panel of Fig.~\ref{fig:spuriousconstrib}. We also give our result for the violation of the angular
condition, which can  be quantified by the sum of the spurious form factors $B_5$ and $B_7$ (cf. Eq.~(\ref{eq:angularconditionviolation})). The result is plotted in the middle panel of Fig.~\ref{fig:spuriousconstrib}. Finally, we have also calculated the spurious form factor $B_6(Q^2)$ from  Eq.~(\ref{eq:GMB6}), which is shown in the bottom panel of Fig.~\ref{fig:spuriousconstrib}. All three cases demonstrate that the spurious contributions cannot be neglected compared to the physical ones and they can contribute significantly  to the current matrix elements. Moreover, they are larger for strongly-bound systems, such as the $\rho$-meson, and less important in weakly-bound systems, such as the deuteron~\cite{Biernat:2011mp}. Therefore, their separation is crucial for the extraction of meaningful physical form factors within our approach (and also within other approaches which violate the angular condition).
\subsection{Comparisons}
In order to make sensible comparisons with other approaches to $\rho$-meson form factors, we look particularly at calculations that use the Gaussian form~(\ref{eq:u00}) for the $\rho$-meson wave function. These are light-front calculations along the lines of Refs.~\cite{Chung:1988my,Carbonell:1998rj,Jaus:2002sv,Choi:2004ww} which differ mainly in the way how the angular condition is dealt with.\footnote{Numerical results for the $\rho$-meson form factors calculated with the harmonic-oscillator wave function~(\ref{eq:u00}) along the lines of Refs.~\cite{Chung:1988my,Carbonell:1998rj} can be found in Ref.~\cite{Jaus:2002sv}.}   In Ref.~\cite{Chung:1988my} no attempt is made to satisfy the angular condition. References~\cite{Carbonell:1998rj,Jaus:2002sv} are both based on the covariant light-front scheme, the difference being that zero modes are, as additional ingredients, taken into account in Ref.~\cite{Jaus:2002sv}. Reference~\cite{Choi:2004ww}, on the other hand, takes into account zero modes, but does 
not use the covariant light-front approach with its spurious contributions.

For the quantitative comparison we adopt the values for the 2 parameters $m_q$ and $a$ of each approach and use them in our calculation.
The predictions for the magnetic dipole moment $\mu_\rho$ and the electric quadrupole moment $Q_\rho$ are compared in Tab.~\ref{tab:rhomuQ}.
\begin{table} [t!]
\begin{center}
\begin{tabular}
{|l|c|c|c|c|}\hline
Ref.&$m_q$ (GeV)&$a$ (GeV)&$\mu_\rho$&$Q_\rho $ \\ \hline
this work&0.34&0.312&2.20&-0.47\\\hline
Choi et al.~\cite{Choi:2004ww} &0.22&0.3659&1.92&-0.43\\
this work&0.22&0.3659&2.33&-0.33\\ \hline
Jaus~\cite{Jaus:2002sv}&0.25&0.28&1.83&-0.33\\
this work&0.25&0.28&2.25&-0.33\\ \hline
Carbonell et al.~\cite{Carbonell:1998rj}&0.25&0.262&2.23& -0.005\\
this work&0.25&0.262&2.231&-0.0058\\\hline
Chung et al.~\cite{Chung:1988my}&0.25&0.316&2.23& -0.19\\
this work &0.25&0.316& 2.27344&-0.253915 \\\hline
\end{tabular}             \end{center}\caption{\label{tab:rhomuQ}
Comparison of the magnetic dipole moment
$\mu_\rho$ (in units $|\,e\,|/2m_{\rho}$) and the electric quadrupole moment $Q_\rho$ (in units $|\,e\,|/m_{\rho}^2$) from different approaches using a harmonic-oscillator confining potential.
}\end{table}
It turns out that our results for $\mu_\rho$ and $Q_\rho$ agree with the calculation \`a la Carbonell et al.~\cite{Carbonell:1998rj}. This is not surprising due to the resemblances between both approaches. Our value for $Q_\rho$ also coincides with the the one from Jaus, Ref.~\cite{Jaus:2002sv}, however the values for $\mu_\rho$ differ significantly. The reason is that current matrix elements which are needed to calculate $\mu_\rho$ are affected by zero-mode contributions, whereas $Q_\rho$ is dominated by $F_2(0)$ for which zero-modes do not play a role. Considerable deviations from the results in Ref.~\cite{Choi:2004ww} are observed for both, the magnetic dipole moment and the electric quadrupole moment. These authors account for zero modes, but do not employ the manifestly covariant light-front approach of Ref.~\cite{Carbonell:1998rj}. By comparison of our results
with the ones from Chung et al., Ref.~\cite{Chung:1988my}, we find that their value for $\mu_\rho$ is about $0.0434$ units of $|\,e\,|/2m_{\rho}$ smaller than ours. This discrepancy is perfectly understood, as it is just the spurious contribution $-B_6(Q^2)+[B_5(Q^2)+B_7(Q^2)]/(1+\eta)$ which admixes to the magnetic form factor when using the usual light-front prescription of Ref.~\cite{Chung:1988my} without separating spurious contributions (see also Ref.~\cite{Carbonell:1998rj}).

In Tab.~\ref{tab:rhomuQpred} we give our predictions for $\mu_\rho$ and $Q_\rho$ using our own parameter values (see Sec.~\ref{subsec:wf}) and compare with results from different sources.
\begin{table}[t!]
\begin{center}
\begin{tabular}
{|l|c|c|}\hline
Ref.&$\mu_\rho$&$Q_\rho$\\ \hline
this work&2.20&-0.47\\
Bagdasaryan et al.~\cite{Bagdasaryan:1984kz}&2.30&-0.45\\
Samsonov~\cite{Samsonov:2003hs}&2.00$\pm$0.3&-\\
Aliev et al.~\cite{Aliev:2004uj}&2.30&-\\
Cardarelli et al.~\cite{Cardarelli:1994yq}&2.23&-0.61\\
Bhagwat et al.~\cite{Bhagwat:2006pu}&2.01&-0.41\\
Hawes et al.~\cite{Hawes:1998bz}&2.69&-0.84\\
De Melo et al.~\cite{deMelo:1997hh}&2.14&-0.79\\
QCDSF~\cite{Gurtler:2008zz}&1.7(3)&-0.015(4)\\
Garcia Gudino et al.~\cite{Gudino:2013jaa}&$2.1\pm0.5$&-\\ \hline
\end{tabular}             \end{center}\caption{\label{tab:rhomuQpred}
Predictions of the magnetic dipole moment
$\mu_\rho$ (in units $|\,e\,|/2m_{\rho}$) and the electric quadrupole moment $Q_\rho$ (in units $|\,e\,|/m_{\rho}^2$) coming from different approaches.
}\end{table}
Our predictions for $\mu_\rho$ and $Q_\rho$ lie within the realm of values obtained by others. Our magnetic moment agrees, in particular, with a recent analysis of Babar data~\cite{Gudino:2013jaa}.

In Tab.~\ref{tab:ffsrho} our form factor results for finite $Q^2$ are confronted with those of Choi et al., Ref.~\cite{Choi:2004ww}, who also use a harmonic-oscillator confining potential. For the purpose of comparison we have again taken the same parameters as in Ref.~\cite{Choi:2004ww}. For $Q^2 \geq 1$~GeV$^2$ both calculations provide comparable results with the largest discrepancies being observed for the charge form factor $G_C$.
\begin{table}
\begin{center}
\begin{tabular}
{|c|c|c|c|}
\hline
$Q^2(\mathrm{GeV}^2)$& &\cite{Choi:2004ww}&this work \\ \hline
&$G_C$&0.38&0.29\\
$Q^2=1$&$G_M$&0.93&0.93\\
&$G_Q$&-0.23&-0.21\\\hline
&$G_C$&0.18&0.12\\
$Q^2=2$&$G_M$&0.59&0.58\\
&$G_Q$&-0.15&-0.14\\\hline
&$G_C$&0.08&0.05\\
$Q^2=3$&$G_M$&0.41&0.41\\
&$G_Q$&-0.10&-0.10\\\hline
\end{tabular}                                             \end{center}
\caption{\label{tab:ffsrho}
Comparison of $\rho$-meson form factors obtained within our approach and the one of Choi et al., Ref.~\cite{Choi:2004ww}, for some fixed values of $Q^2$.
Both calculations were done with the same harmonic-oscillator model with the parameters $m_q=0.22~\mathrm{GeV}$ and $a=0.3659~\mathrm{GeV}$.
}\end{table}

In Tab.~\ref{tab:ffsrhopred} we finally compare our predictions for the $\rho$-meson form factors, that we have already plotted in Sect.~\ref{subsec:prediction}, with those of QCD sum rules~\cite{Aliev:2004uj,Braguta:2004kx} and Bethe-Salpeter-Dyson-Schwinger methods~\cite{Bhagwat:2006pu,Hawes:1998bz}.
\begin{table}
\begin{center}
\begin{tabular}
{|c|c|c|c|c|c|c|}
\hline
$Q^2(\mathrm{GeV}^2)$& &this work&\cite{Bhagwat:2006pu}&\cite{Hawes:1998bz}&\cite{Aliev:2004uj}&\cite{Braguta:2004kx}\\ \hline
&$G_C$&0.29&0.22&0.17&0.25&0.10\\
$Q^2=1$&$G_M$&0.85&0.57&0.85&0.58&0.46\\
&$G_Q$&-0.21&-0.11&-0.51&-0.49&-0.16\\\hline
&$G_C$&0.11&0.08&0.04&0.13&0.16\\
$Q^2=2$&$G_M$&0.47&0.27&0.45&0.28&0.27\\
&$G_Q$&-0.12&-0.05&-0.32&-0.24&-0.11\\\hline
&$G_C$&0.05&&0.11&0.08&-0.03\\
$Q^2=3$&$G_M$&0.30&&0.25&0.17&0.18\\
&$G_Q$&-0.07&&-0.23&-0.15&-0.10\\\hline
\end{tabular}                                             \end{center}
\caption{\label{tab:ffsrhopred}
The $\rho$-meson form factors obtained with our own parametrization in comparison with predictions from QCD sum rules~\cite{Aliev:2004uj,Braguta:2004kx} and Bethe-Salpeter-Dyson-Schwinger methods~\cite{Bhagwat:2006pu,Hawes:1998bz}. }
\end{table}
We observe that our electric charge form factor $G_C$ for $Q^2 = 1$~GeV$^2$ lies above the predictions of these other approaches. For the higher $Q^2$ it is somewhere between the values of the other approaches. Our magnetic form factor $G_M$ lies above the values of the other approaches in the whole range of $Q^2$, whereas our electric quadrupole form factor $G_Q$ is within the range of sum-rule and Bethe-Salpeter-Dyson-Schwinger results. To conclude, in view of the simplicity of our harmonic-oscillator model the results for the $\rho$-meson form factors look quite reasonable and fall within the range of other model predictions. Experimentally little is known about the electromagnetic $\rho$-meson form factors which could be used to discriminate between different models and
approaches.
\section{Summary and conclusions}\label{sec:summary}
The point form is the least utilized of Dirac's forms of relativistic dynamics, although it has several advantageous features. Its key benefit is the property that Lorentz transformations are kinematic and only space-time translations are affected by interactions. This natural way of separating the kinematic from the dynamic generators allows for a manifest Lorentz covariant formulation of operator equations and yields a simple behavior of wave functions under Lorentz transformations. In the present work we have exploited the virtues of point-form relativistic quantum mechanics to analyze the electromagnetic structure of $q\bar q$ mesons within constituent-quark models. Applying the coupled-channel formalism developed earlier~\cite{Biernat:2009my} we have derived an expression for the electromagnetic $\rho$-meson current in terms of quark currents and the $\rho$-meson wave function. Our current is hermitian and transforms correctly under Lorentz transformations. Its Lorentz structure is, however, not 
completely determined by the incoming and outgoing meson momenta and spins. It turned out that additional, unphysical (or spurious) Lorentz-covariant terms are necessary to parameterize the entire current. These spurious contributions depend on the electron momenta and they are a consequence of the violation of cluster separability in the Bakamjian-Thomas framework, which we used to ensure Poincar\'e invariance. Furthermore, both the physical and the spurious form factors (associated with the spurious contributions) depend on the 2 independent Lorentz invariants that can be constructed from the incoming and outgoing meson and electron momenta, i.e. Mandelstam $t$, the four-momentum transfer squared, and also Mandelstam $s$, the total invariant mass squared of the electron-meson system.

For pseudoscalar mesons such as the pion, the current derived along the same lines is conserved and can be parametrized by 1 physical and 1 spurious form factor~\cite{Biernat:2007dn}. The structure of this current reveals an interesting correspondence to the covariant light-front approach~\cite{Karmanov:1994ck,Carbonell:1998rj}. The electromagnetic pion current in the covariant light-front formalism contains also a spurious contribution, which is associated with an arbitrary light-like four-vector $\omega^\mu$ that describes the orientation of the light front. The dependence on this light-front orientation is a consequence of demanding manifest Lorentz covariance in the light-front approach. In our case,  the spurious contribution could be removed from the pion current by taking the limit Mandelstam $s\rightarrow\infty$. This limit abolishes also the unwanted $s$-dependence of the pion form factor. The resulting analytical expression is equivalent to the one that is extracted from the plus component of a 
spectator current in usual light-front dynamics in the $q^+= 0$ frame~\cite{Chung:1988mu}. As in the pseudoscalar case, the Lorentz structure of our vector-meson current, which contains 3 physical and 8 spurious contributions, resembled the corresponding current of the covariant light-front formalism. However, unlike the pion case, by taking the limit $s\rightarrow\infty$, only 4 spurious contributions, proportional to $1/\sqrt{s}$, were removed. The remaining 4 spurious contributions violate current conservation and the angular condition. This parallels again the situation of the covariant light-front approach after adopting the common choice $\omega=(1,0,0,-1)$. The 4 spurious contributions that are proportional to $\omega^\mu$ vanish in this case for the plus component of the spin-1 spectator current from which the form factors are usually extracted. As in our approach the surviving 4 spurious contributions violate current conservation and the angular condition. By means of the projection technique 
proposed by Karmanov and Smirnov~\cite{Karmanov:1994ck} it is, however, possible to neatly separate the physical from the unphysical contributions. The same projection technique is also applicable in our case. Alternatively we can exploit the observation that one can find 3 independent current matrix elements which are (to leading order in a $1/\sqrt{s}$ expansion) free of spurious contributions and can thus be taken to extract the physical form factors directly. 

Our numerical studies for the $\rho$-meson magnetic dipole moment and electric quadrupole moment with a simple harmonic-oscillator wave function showed agreement with the corresponding results obtained from the covariant light-front prescription. The results for the magnetic dipole moment obtained from the usual light-front prescription  differed from our results precisely by the value of the spurious contribution that is ignored when extracting the magnetic form factor within the usual (non-covariant) light-front approach.  It seems to us quite remarkable that point-form and (covariant) light-front dynamics give the same results. This is what one would expect for physical reasons without approximations. Here, however, we are dealing with simple spectator currents in both approaches which lead to unphysical contributions in the currents. These unphysical contributions are of different origin, namely wrong cluster properties in our approach and an unwanted dependence on the light-front orientation in the  
covariant light-front approach. Nevertheless, after getting rid of the spurious contributions we end up with the same current. As a byproduct the size of the spurious contributions gives us a measure for the violation of cluster separability, which turned out to be by far non negligible in strongly bound systems.

With the present study of electromagnetic vector-meson form factors we have extended the scope of the employed relativistic coupled-channel formalism.  In foregoing work this formalism has been applied to investigate the electroweak properties of heavy-light mesons~\cite{GomezRocha:2012zd} and to prove that the correct heavy-quark-symmetry properties emerge in the heavy-quark limit. Furthermore, the form factors of the deuteron have been calculated within a Walecka-type model~\cite{Bakker:2010,ref.01} for the NN interaction with instantaneous~\cite{Biernat:2011mp} and also dynamical~\cite{Gomez-Rocha:2013bga} $\sigma$ and $\omega$-meson exchanges.  Dynamical particle exchange between the bound-state constituents gives rise to exchange-currents. These can also be accommodated within our coupled channel 
framework. Ongoing studies, e.g., deal with the effect of dynamical pion exchange between constituent quarks on the electromagnetic nucleon form factors~\cite{Kupelwieser:2013nqa}. Non-valence components in hadrons can also be treated and their role in hadron decay form factors is a further subject of investigations~\cite{Gomez-Rocha:2013zma}.

\begin{acknowledgements}
This work received financial support from the \lq\lq Fonds zur F\"{o}rderung der wissenschaftlichen Forschung in \"{O}sterreich'' under grant No. FWF DK W1203-N16, as well as from the Province of Styria, Austria under a PhD grant. This work was also partially supported by the \lq\lq Funda\c c\~ao para a Ci\^encia e a Tecnologia (FCT)'' under grant Nos.~PTDC/FIS/113940/2009 and CFTP-FCT (PEst-OE/FIS/U/0777/2013), and by the European Union under the HadronPhysics3 Grant No. 283286.
\end{acknowledgements}

\bibliographystyle{apsrev4-1}
\bibliography{Bib-v3}

\begin{thebibliography}{54}%
\makeatletter
\providecommand \@ifxundefined [1]{%
 \@ifx{#1\undefined}
}%
\providecommand \@ifnum [1]{%
 \ifnum #1\expandafter \@firstoftwo
 \else \expandafter \@secondoftwo
 \fi
}%
\providecommand \@ifx [1]{%
 \ifx #1\expandafter \@firstoftwo
 \else \expandafter \@secondoftwo
 \fi
}%
\providecommand \natexlab [1]{#1}%
\providecommand \enquote  [1]{``#1''}%
\providecommand \bibnamefont  [1]{#1}%
\providecommand \bibfnamefont [1]{#1}%
\providecommand \citenamefont [1]{#1}%
\providecommand \href@noop [0]{\@secondoftwo}%
\providecommand \href [0]{\begingroup \@sanitize@url \@href}%
\providecommand \@href[1]{\@@startlink{#1}\@@href}%
\providecommand \@@href[1]{\endgroup#1\@@endlink}%
\providecommand \@sanitize@url [0]{\catcode `\\12\catcode `\$12\catcode
  `\&12\catcode `\#12\catcode `\^12\catcode `\_12\catcode `\%12\relax}%
\providecommand \@@startlink[1]{}%
\providecommand \@@endlink[0]{}%
\providecommand \url  [0]{\begingroup\@sanitize@url \@url }%
\providecommand \@url [1]{\endgroup\@href {#1}{\urlprefix }}%
\providecommand \urlprefix  [0]{URL }%
\providecommand \Eprint [0]{\href }%
\providecommand \doibase [0]{http://dx.doi.org/}%
\providecommand \selectlanguage [0]{\@gobble}%
\providecommand \bibinfo  [0]{\@secondoftwo}%
\providecommand \bibfield  [0]{\@secondoftwo}%
\providecommand \translation [1]{[#1]}%
\providecommand \BibitemOpen [0]{}%
\providecommand \bibitemStop [0]{}%
\providecommand \bibitemNoStop [0]{.\EOS\space}%
\providecommand \EOS [0]{\spacefactor3000\relax}%
\providecommand \BibitemShut  [1]{\csname bibitem#1\endcsname}%
\let\auto@bib@innerbib\@empty
\bibitem [{\citenamefont {Lev}(1995)}]{Lev:1994au}%
  \BibitemOpen
  \bibfield  {author} {\bibinfo {author} {\bibfnamefont {F.~M.}\ \bibnamefont
  {Lev}},\ }\href@noop {} {\bibfield  {journal} {\bibinfo  {journal} {Annals
  Phys.}\ }\textbf {\bibinfo {volume} {237}},\ \bibinfo {pages} {355} (\bibinfo
  {year} {1995})}\BibitemShut {NoStop}%
\bibitem [{\citenamefont {Bakamjian}\ and\ \citenamefont
  {Thomas}(1953)}]{Bakamjian:1953kh}%
  \BibitemOpen
  \bibfield  {author} {\bibinfo {author} {\bibfnamefont {B.}~\bibnamefont
  {Bakamjian}}\ and\ \bibinfo {author} {\bibfnamefont {L.~H.}\ \bibnamefont
  {Thomas}},\ }\href {\doibase 10.1103/PhysRev.92.1300} {\bibfield  {journal}
  {\bibinfo  {journal} {Phys. Rev.}\ }\textbf {\bibinfo {volume} {92}},\
  \bibinfo {pages} {1300} (\bibinfo {year} {1953})}\BibitemShut {NoStop}%
\bibitem [{\citenamefont {Dirac}(1949)}]{Dirac:1949cp}%
  \BibitemOpen
  \bibfield  {author} {\bibinfo {author} {\bibfnamefont {P.~A.~M.}\
  \bibnamefont {Dirac}},\ }\href@noop {} {\bibfield  {journal} {\bibinfo
  {journal} {Rev. Mod. Phys.}\ }\textbf {\bibinfo {volume} {21}},\ \bibinfo
  {pages} {392} (\bibinfo {year} {1949})}\BibitemShut {NoStop}%
\bibitem [{\citenamefont {Sokolov}\ and\ \citenamefont
  {Shatnyi}(1979)}]{Sokolov:1977im}%
  \BibitemOpen
  \bibfield  {author} {\bibinfo {author} {\bibfnamefont {S.~N.}\ \bibnamefont
  {Sokolov}}\ and\ \bibinfo {author} {\bibfnamefont {A.~N.}\ \bibnamefont
  {Shatnyi}},\ }\href {\doibase 10.1007/BF01018583} {\bibfield  {journal}
  {\bibinfo  {journal} {Theor. Math. Phys.}\ }\textbf {\bibinfo {volume}
  {37}},\ \bibinfo {pages} {1029} (\bibinfo {year} {1979})}\BibitemShut
  {NoStop}%
\bibitem [{\citenamefont {Biernat}\ \emph {et~al.}(2011)\citenamefont
  {Biernat}, \citenamefont {Klink},\ and\ \citenamefont
  {Schweiger}}]{Biernat:2010tp}%
  \BibitemOpen
  \bibfield  {author} {\bibinfo {author} {\bibfnamefont {E.~P.}\ \bibnamefont
  {Biernat}}, \bibinfo {author} {\bibfnamefont {W.~H.}\ \bibnamefont {Klink}},
  \ and\ \bibinfo {author} {\bibfnamefont {W.}~\bibnamefont {Schweiger}},\
  }\href@noop {} {\bibfield  {journal} {\bibinfo  {journal} {Few Body Syst.}\
  }\textbf {\bibinfo {volume} {49}},\ \bibinfo {pages} {149} (\bibinfo {year}
  {2011})}\BibitemShut {NoStop}%
\bibitem [{\citenamefont {Klink}(1998{\natexlab{a}})}]{Klink:1998zz}%
  \BibitemOpen
  \bibfield  {author} {\bibinfo {author} {\bibfnamefont {W.~H.}\ \bibnamefont
  {Klink}},\ }\href {\doibase 10.1103/PhysRevC.58.3617} {\bibfield  {journal}
  {\bibinfo  {journal} {Phys. Rev.}\ }\textbf {\bibinfo {volume} {C58}},\
  \bibinfo {pages} {3617} (\bibinfo {year} {1998}{\natexlab{a}})}\BibitemShut
  {NoStop}%
\bibitem [{\citenamefont {Klink}(1998{\natexlab{b}})}]{Klink:1998qf}%
  \BibitemOpen
  \bibfield  {author} {\bibinfo {author} {\bibfnamefont {W.~H.}\ \bibnamefont
  {Klink}},\ }\href@noop {} {\bibfield  {journal} {\bibinfo  {journal} {Phys.
  Rev.}\ }\textbf {\bibinfo {volume} {C58}},\ \bibinfo {pages} {3587} (\bibinfo
  {year} {1998}{\natexlab{b}})}\BibitemShut {NoStop}%
\bibitem [{\citenamefont {Allen}\ and\ \citenamefont
  {Klink}(1998)}]{Allen:1998hb}%
  \BibitemOpen
  \bibfield  {author} {\bibinfo {author} {\bibfnamefont {T.~W.}\ \bibnamefont
  {Allen}}\ and\ \bibinfo {author} {\bibfnamefont {W.~H.}\ \bibnamefont
  {Klink}},\ }\href {\doibase 10.1103/PhysRevC.58.3670} {\bibfield  {journal}
  {\bibinfo  {journal} {Phys. Rev.}\ }\textbf {\bibinfo {volume} {C58}},\
  \bibinfo {pages} {3670} (\bibinfo {year} {1998})}\BibitemShut {NoStop}%
\bibitem [{\citenamefont {Allen}\ \emph {et~al.}(2001)\citenamefont {Allen},
  \citenamefont {Klink},\ and\ \citenamefont {Polyzou}}]{Allen:2000ge}%
  \BibitemOpen
  \bibfield  {author} {\bibinfo {author} {\bibfnamefont {T.~W.}\ \bibnamefont
  {Allen}}, \bibinfo {author} {\bibfnamefont {W.~H.}\ \bibnamefont {Klink}}, \
  and\ \bibinfo {author} {\bibfnamefont {W.~N.}\ \bibnamefont {Polyzou}},\
  }\href {\doibase 10.1103/PhysRevC.63.034002} {\bibfield  {journal} {\bibinfo
  {journal} {Phys. Rev.}\ }\textbf {\bibinfo {volume} {C63}},\ \bibinfo {pages}
  {034002} (\bibinfo {year} {2001})}\BibitemShut {NoStop}%
\bibitem [{\citenamefont {Wagenbrunn}\ \emph {et~al.}(2001)\citenamefont
  {Wagenbrunn}, \citenamefont {Boffi}, \citenamefont {Klink}, \citenamefont
  {Plessas},\ and\ \citenamefont {Radici}}]{Wagenbrunn:2000es}%
  \BibitemOpen
  \bibfield  {author} {\bibinfo {author} {\bibfnamefont {R.~F.}\ \bibnamefont
  {Wagenbrunn}}, \bibinfo {author} {\bibfnamefont {S.}~\bibnamefont {Boffi}},
  \bibinfo {author} {\bibfnamefont {W.}~\bibnamefont {Klink}}, \bibinfo
  {author} {\bibfnamefont {W.}~\bibnamefont {Plessas}}, \ and\ \bibinfo
  {author} {\bibfnamefont {M.}~\bibnamefont {Radici}},\ }\href {\doibase
  10.1016/S0370-2693(01)00622-0} {\bibfield  {journal} {\bibinfo  {journal}
  {Phys. Lett.}\ }\textbf {\bibinfo {volume} {B511}},\ \bibinfo {pages} {33}
  (\bibinfo {year} {2001})}\BibitemShut {NoStop}%
\bibitem [{\citenamefont {Boffi}\ \emph {et~al.}(2002)\citenamefont {Boffi}
  \emph {et~al.}}]{Boffi:2001zb}%
  \BibitemOpen
  \bibfield  {author} {\bibinfo {author} {\bibfnamefont {S.}~\bibnamefont
  {Boffi}} \emph {et~al.},\ }\href {\doibase 10.1007/s10050-002-8784-3}
  {\bibfield  {journal} {\bibinfo  {journal} {Eur. Phys. J.}\ }\textbf
  {\bibinfo {volume} {A14}},\ \bibinfo {pages} {17} (\bibinfo {year}
  {2002})}\BibitemShut {NoStop}%
\bibitem [{\citenamefont {Melde}\ \emph {et~al.}(2007)\citenamefont {Melde},
  \citenamefont {Berger}, \citenamefont {Canton}, \citenamefont {Plessas},\
  and\ \citenamefont {Wagenbrunn}}]{Melde:2007zz}%
  \BibitemOpen
  \bibfield  {author} {\bibinfo {author} {\bibfnamefont {T.}~\bibnamefont
  {Melde}}, \bibinfo {author} {\bibfnamefont {K.}~\bibnamefont {Berger}},
  \bibinfo {author} {\bibfnamefont {L.}~\bibnamefont {Canton}}, \bibinfo
  {author} {\bibfnamefont {W.}~\bibnamefont {Plessas}}, \ and\ \bibinfo
  {author} {\bibfnamefont {R.~F.}\ \bibnamefont {Wagenbrunn}},\ }\href
  {\doibase 10.1103/PhysRevD.76.074020} {\bibfield  {journal} {\bibinfo
  {journal} {Phys. Rev.}\ }\textbf {\bibinfo {volume} {D76}},\ \bibinfo {pages}
  {074020} (\bibinfo {year} {2007})}\BibitemShut {NoStop}%
\bibitem [{\citenamefont {Klink}(2003)}]{Klink:2000pp}%
  \BibitemOpen
  \bibfield  {author} {\bibinfo {author} {\bibfnamefont {W.~H.}\ \bibnamefont
  {Klink}},\ }\href@noop {} {\bibfield  {journal} {\bibinfo  {journal} {Nucl.
  Phys.}\ }\textbf {\bibinfo {volume} {A716}},\ \bibinfo {pages} {123}
  (\bibinfo {year} {2003})}\BibitemShut {NoStop}%
\bibitem [{\citenamefont {Biernat}\ \emph {et~al.}(2009)\citenamefont
  {Biernat}, \citenamefont {Schweiger}, \citenamefont {Fuchsberger},\ and\
  \citenamefont {Klink}}]{Biernat:2009my}%
  \BibitemOpen
  \bibfield  {author} {\bibinfo {author} {\bibfnamefont {E.~P.}\ \bibnamefont
  {Biernat}}, \bibinfo {author} {\bibfnamefont {W.}~\bibnamefont {Schweiger}},
  \bibinfo {author} {\bibfnamefont {K.}~\bibnamefont {Fuchsberger}}, \ and\
  \bibinfo {author} {\bibfnamefont {W.~H.}\ \bibnamefont {Klink}},\ }\href@noop
  {} {\bibfield  {journal} {\bibinfo  {journal} {Phys. Rev.}\ }\textbf
  {\bibinfo {volume} {C79}},\ \bibinfo {pages} {055203} (\bibinfo {year}
  {2009})}\BibitemShut {NoStop}%
\bibitem [{\citenamefont {Gomez-Rocha}\ and\ \citenamefont
  {Schweiger}(2012)}]{GomezRocha:2012zd}%
  \BibitemOpen
  \bibfield  {author} {\bibinfo {author} {\bibfnamefont {M.}~\bibnamefont
  {Gomez-Rocha}}\ and\ \bibinfo {author} {\bibfnamefont {W.}~\bibnamefont
  {Schweiger}},\ }\href@noop {} {\bibfield  {journal} {\bibinfo  {journal}
  {Phys.Rev.}\ }\textbf {\bibinfo {volume} {D86}},\ \bibinfo {pages} {053010}
  (\bibinfo {year} {2012})}\BibitemShut {NoStop}%
\bibitem [{\citenamefont {Sokolov}(1979)}]{Sokolov:1977ym}%
  \BibitemOpen
  \bibfield  {author} {\bibinfo {author} {\bibfnamefont {S.~N.}\ \bibnamefont
  {Sokolov}},\ }\href {\doibase 10.1007/BF01036481} {\bibfield  {journal}
  {\bibinfo  {journal} {Theor. Math. Phys.}\ }\textbf {\bibinfo {volume}
  {36}},\ \bibinfo {pages} {682} (\bibinfo {year} {1979})}\BibitemShut
  {NoStop}%
\bibitem [{\citenamefont {Coester}\ and\ \citenamefont
  {Polyzou}(1982)}]{Coester:1982vt}%
  \BibitemOpen
  \bibfield  {author} {\bibinfo {author} {\bibfnamefont {F.}~\bibnamefont
  {Coester}}\ and\ \bibinfo {author} {\bibfnamefont {W.~N.}\ \bibnamefont
  {Polyzou}},\ }\href {\doibase 10.1103/PhysRevD.26.1348} {\bibfield  {journal}
  {\bibinfo  {journal} {Phys. Rev.}\ }\textbf {\bibinfo {volume} {D26}},\
  \bibinfo {pages} {1348} (\bibinfo {year} {1982})}\BibitemShut {NoStop}%
\bibitem [{\citenamefont {Keister}\ and\ \citenamefont
  {Polyzou}(1991)}]{Keister:1991sb}%
  \BibitemOpen
  \bibfield  {author} {\bibinfo {author} {\bibfnamefont {B.~D.}\ \bibnamefont
  {Keister}}\ and\ \bibinfo {author} {\bibfnamefont {W.~N.}\ \bibnamefont
  {Polyzou}},\ }\href@noop {} {\bibfield  {journal} {\bibinfo  {journal} {Adv.
  Nucl. Phys.}\ }\textbf {\bibinfo {volume} {20}},\ \bibinfo {pages} {225}
  (\bibinfo {year} {1991})}\BibitemShut {NoStop}%
\bibitem [{\citenamefont {Karmanov}\ and\ \citenamefont
  {Smirnov}(1994)}]{Karmanov:1994ck}%
  \BibitemOpen
  \bibfield  {author} {\bibinfo {author} {\bibfnamefont {V.~A.}\ \bibnamefont
  {Karmanov}}\ and\ \bibinfo {author} {\bibfnamefont {A.~V.}\ \bibnamefont
  {Smirnov}},\ }\href {\doibase 10.1016/0375-9474(94)90374-3} {\bibfield
  {journal} {\bibinfo  {journal} {Nucl. Phys.}\ }\textbf {\bibinfo {volume}
  {A575}},\ \bibinfo {pages} {520} (\bibinfo {year} {1994})}\BibitemShut
  {NoStop}%
\bibitem [{\citenamefont {Carbonell}\ \emph {et~al.}(1998)\citenamefont
  {Carbonell}, \citenamefont {Desplanques}, \citenamefont {Karmanov},\ and\
  \citenamefont {Mathiot}}]{Carbonell:1998rj}%
  \BibitemOpen
  \bibfield  {author} {\bibinfo {author} {\bibfnamefont {J.}~\bibnamefont
  {Carbonell}}, \bibinfo {author} {\bibfnamefont {B.}~\bibnamefont
  {Desplanques}}, \bibinfo {author} {\bibfnamefont {V.~A.}\ \bibnamefont
  {Karmanov}}, \ and\ \bibinfo {author} {\bibfnamefont {J.~F.}\ \bibnamefont
  {Mathiot}},\ }\href {\doibase 10.1016/S0370-1573(97)00090-2} {\bibfield
  {journal} {\bibinfo  {journal} {Phys. Rept.}\ }\textbf {\bibinfo {volume}
  {300}},\ \bibinfo {pages} {215} (\bibinfo {year} {1998})}\BibitemShut
  {NoStop}%
\bibitem [{\citenamefont {Coester}(1965)}]{Coester:1965zz}%
  \BibitemOpen
  \bibfield  {author} {\bibinfo {author} {\bibfnamefont {F.}~\bibnamefont
  {Coester}},\ }\href@noop {} {\bibfield  {journal} {\bibinfo  {journal} {Helv.
  Phys. Acta}\ }\textbf {\bibinfo {volume} {38}},\ \bibinfo {pages} {7}
  (\bibinfo {year} {1965})}\BibitemShut {NoStop}%
\bibitem [{\citenamefont {Keister}\ and\ \citenamefont
  {Polyzou}(2012)}]{PhysRevC.86.014002}%
  \BibitemOpen
  \bibfield  {author} {\bibinfo {author} {\bibfnamefont {B.~D.}\ \bibnamefont
  {Keister}}\ and\ \bibinfo {author} {\bibfnamefont {W.~N.}\ \bibnamefont
  {Polyzou}},\ }\href {\doibase 10.1103/PhysRevC.86.014002} {\bibfield
  {journal} {\bibinfo  {journal} {Phys. Rev. C}\ }\textbf {\bibinfo {volume}
  {86}},\ \bibinfo {pages} {014002} (\bibinfo {year} {2012})}\BibitemShut
  {NoStop}%
\bibitem [{\citenamefont {Biernat}(2011)}]{Biernat:2011mp}%
  \BibitemOpen
  \bibfield  {author} {\bibinfo {author} {\bibfnamefont {E.~P.}\ \bibnamefont
  {Biernat}},\ }\href@noop {} {Ph.D. thesis},\ \bibinfo  {school}
  {Karl-Franzens University of Graz} (\bibinfo {year} {2011}),\ \Eprint
  {http://arxiv.org/abs/1110.3180} {arXiv:1110.3180} \BibitemShut {NoStop}%
\bibitem [{\citenamefont {Gomez-Rocha}(2013)}]{Gomez-Rocha:2013bga}%
  \BibitemOpen
  \bibfield  {author} {\bibinfo {author} {\bibfnamefont {M.}~\bibnamefont
  {Gomez-Rocha}},\ }\href@noop {} {Ph.D. thesis},\ \bibinfo  {school}
  {Karl-Franzens University of Graz} (\bibinfo {year} {2013}),\ \Eprint
  {http://arxiv.org/abs/1306.1248} {arXiv:1306.1248} \BibitemShut {NoStop}%
\bibitem [{\citenamefont {Krassnigg}\ \emph {et~al.}(2003)\citenamefont
  {Krassnigg}, \citenamefont {Schweiger},\ and\ \citenamefont
  {Klink}}]{Krassnigg:2003gh}%
  \BibitemOpen
  \bibfield  {author} {\bibinfo {author} {\bibfnamefont {A.}~\bibnamefont
  {Krassnigg}}, \bibinfo {author} {\bibfnamefont {W.}~\bibnamefont
  {Schweiger}}, \ and\ \bibinfo {author} {\bibfnamefont {W.~H.}\ \bibnamefont
  {Klink}},\ }\href {\doibase 10.1103/PhysRevC.67.064003} {\bibfield  {journal}
  {\bibinfo  {journal} {Phys. Rev.}\ }\textbf {\bibinfo {volume} {C67}},\
  \bibinfo {pages} {064003} (\bibinfo {year} {2003})}\BibitemShut {NoStop}%
\bibitem [{\citenamefont {Chung}\ \emph
  {et~al.}(1988{\natexlab{a}})\citenamefont {Chung}, \citenamefont {Coester},\
  and\ \citenamefont {Polyzou}}]{Chung:1988mu}%
  \BibitemOpen
  \bibfield  {author} {\bibinfo {author} {\bibfnamefont {P.~L.}\ \bibnamefont
  {Chung}}, \bibinfo {author} {\bibfnamefont {F.}~\bibnamefont {Coester}}, \
  and\ \bibinfo {author} {\bibfnamefont {W.~N.}\ \bibnamefont {Polyzou}},\
  }\href {\doibase 10.1016/0370-2693(88)90995-1} {\bibfield  {journal}
  {\bibinfo  {journal} {Phys. Lett.}\ }\textbf {\bibinfo {volume} {B205}},\
  \bibinfo {pages} {545} (\bibinfo {year} {1988}{\natexlab{a}})}\BibitemShut
  {NoStop}%
\bibitem [{\citenamefont {Simula}(2002)}]{Simula:2002vm}%
  \BibitemOpen
  \bibfield  {author} {\bibinfo {author} {\bibfnamefont {S.}~\bibnamefont
  {Simula}},\ }\href {\doibase 10.1103/PhysRevC.66.035201} {\bibfield
  {journal} {\bibinfo  {journal} {Phys. Rev.}\ }\textbf {\bibinfo {volume}
  {C66}},\ \bibinfo {pages} {035201} (\bibinfo {year} {2002})}\BibitemShut
  {NoStop}%
\bibitem [{\citenamefont {Buck}\ and\ \citenamefont
  {Gross}(1979)}]{Buck:1979ff}%
  \BibitemOpen
  \bibfield  {author} {\bibinfo {author} {\bibfnamefont {W.~W.}\ \bibnamefont
  {Buck}}\ and\ \bibinfo {author} {\bibfnamefont {F.}~\bibnamefont {Gross}},\
  }\href {\doibase 10.1103/PhysRevD.20.2361} {\bibfield  {journal} {\bibinfo
  {journal} {Phys. Rev.}\ }\textbf {\bibinfo {volume} {D20}},\ \bibinfo {pages}
  {2361} (\bibinfo {year} {1979})}\BibitemShut {NoStop}%
\bibitem [{\citenamefont {Grach}\ and\ \citenamefont
  {Kondratyuk}(1984)}]{Grach:1983hd}%
  \BibitemOpen
  \bibfield  {author} {\bibinfo {author} {\bibfnamefont {I.~L.}\ \bibnamefont
  {Grach}}\ and\ \bibinfo {author} {\bibfnamefont {L.~A.}\ \bibnamefont
  {Kondratyuk}},\ }\href@noop {} {\bibfield  {journal} {\bibinfo  {journal}
  {Sov. J. Nucl. Phys.}\ }\textbf {\bibinfo {volume} {39}},\ \bibinfo {pages}
  {198} (\bibinfo {year} {1984})}\BibitemShut {NoStop}%
\bibitem [{\citenamefont {Bakker}\ and\ \citenamefont
  {Ji}(2002)}]{Bakker:2002aw}%
  \BibitemOpen
  \bibfield  {author} {\bibinfo {author} {\bibfnamefont {B.~L.~G.}\
  \bibnamefont {Bakker}}\ and\ \bibinfo {author} {\bibfnamefont {C.-R.}\
  \bibnamefont {Ji}},\ }\href {\doibase 10.1103/PhysRevD.65.073002} {\bibfield
  {journal} {\bibinfo  {journal} {Phys. Rev.}\ }\textbf {\bibinfo {volume}
  {D65}},\ \bibinfo {pages} {073002} (\bibinfo {year} {2002})}\BibitemShut
  {NoStop}%
\bibitem [{\citenamefont {Chung}\ \emph
  {et~al.}(1988{\natexlab{b}})\citenamefont {Chung}, \citenamefont {Polyzou},
  \citenamefont {Coester},\ and\ \citenamefont {Keister}}]{Chung:1988my}%
  \BibitemOpen
  \bibfield  {author} {\bibinfo {author} {\bibfnamefont {P.~L.}\ \bibnamefont
  {Chung}}, \bibinfo {author} {\bibfnamefont {W.~N.}\ \bibnamefont {Polyzou}},
  \bibinfo {author} {\bibfnamefont {F.}~\bibnamefont {Coester}}, \ and\
  \bibinfo {author} {\bibfnamefont {B.~D.}\ \bibnamefont {Keister}},\ }\href
  {\doibase 10.1103/PhysRevC.37.2000} {\bibfield  {journal} {\bibinfo
  {journal} {Phys. Rev.}\ }\textbf {\bibinfo {volume} {C37}},\ \bibinfo {pages}
  {2000} (\bibinfo {year} {1988}{\natexlab{b}})}\BibitemShut {NoStop}%
\bibitem [{\citenamefont {Brodsky}\ and\ \citenamefont
  {Hiller}(1992)}]{Brodsky:1992px}%
  \BibitemOpen
  \bibfield  {author} {\bibinfo {author} {\bibfnamefont {S.~J.}\ \bibnamefont
  {Brodsky}}\ and\ \bibinfo {author} {\bibfnamefont {J.~R.}\ \bibnamefont
  {Hiller}},\ }\href {\doibase 10.1103/PhysRevD.46.2141} {\bibfield  {journal}
  {\bibinfo  {journal} {Phys. Rev.}\ }\textbf {\bibinfo {volume} {D46}},\
  \bibinfo {pages} {2141} (\bibinfo {year} {1992})}\BibitemShut {NoStop}%
\bibitem [{\citenamefont {Frankfurt}\ \emph {et~al.}(1993)\citenamefont
  {Frankfurt}, \citenamefont {Strikman},\ and\ \citenamefont
  {Frederico}}]{Frankfurt:1993ut}%
  \BibitemOpen
  \bibfield  {author} {\bibinfo {author} {\bibfnamefont {L.~L.}\ \bibnamefont
  {Frankfurt}}, \bibinfo {author} {\bibfnamefont {M.}~\bibnamefont {Strikman}},
  \ and\ \bibinfo {author} {\bibfnamefont {T.}~\bibnamefont {Frederico}},\
  }\href {\doibase 10.1103/PhysRevC.48.2182} {\bibfield  {journal} {\bibinfo
  {journal} {Phys. Rev.}\ }\textbf {\bibinfo {volume} {C48}},\ \bibinfo {pages}
  {2182} (\bibinfo {year} {1993})}\BibitemShut {NoStop}%
\bibitem [{\citenamefont {Cardarelli}\ \emph {et~al.}(1995)\citenamefont
  {Cardarelli}, \citenamefont {Grach}, \citenamefont {Narodetsky},
  \citenamefont {Salme},\ and\ \citenamefont {Simula}}]{Cardarelli:1994yq}%
  \BibitemOpen
  \bibfield  {author} {\bibinfo {author} {\bibfnamefont {F.}~\bibnamefont
  {Cardarelli}}, \bibinfo {author} {\bibfnamefont {I.~L.}\ \bibnamefont
  {Grach}}, \bibinfo {author} {\bibfnamefont {I.~M.}\ \bibnamefont
  {Narodetsky}}, \bibinfo {author} {\bibfnamefont {G.}~\bibnamefont {Salme}}, \
  and\ \bibinfo {author} {\bibfnamefont {S.}~\bibnamefont {Simula}},\ }\href
  {\doibase 10.1016/0370-2693(95)00230-I} {\bibfield  {journal} {\bibinfo
  {journal} {Phys. Lett.}\ }\textbf {\bibinfo {volume} {B349}},\ \bibinfo
  {pages} {393} (\bibinfo {year} {1995})}\BibitemShut {NoStop}%
\bibitem [{\citenamefont {Karmanov}(1996)}]{Karmanov:1996qc}%
  \BibitemOpen
  \bibfield  {author} {\bibinfo {author} {\bibfnamefont {V.~A.}\ \bibnamefont
  {Karmanov}},\ }\href {\doibase 10.1016/0375-9474(96)00260-6} {\bibfield
  {journal} {\bibinfo  {journal} {Nucl. Phys.}\ }\textbf {\bibinfo {volume}
  {A608}},\ \bibinfo {pages} {316} (\bibinfo {year} {1996})}\BibitemShut
  {NoStop}%
\bibitem [{\citenamefont {Keister}(1994)}]{Keister:1993mg}%
  \BibitemOpen
  \bibfield  {author} {\bibinfo {author} {\bibfnamefont {B.~D.}\ \bibnamefont
  {Keister}},\ }\href {\doibase 10.1103/PhysRevD.49.1500} {\bibfield  {journal}
  {\bibinfo  {journal} {Phys. Rev.}\ }\textbf {\bibinfo {volume} {D49}},\
  \bibinfo {pages} {1500} (\bibinfo {year} {1994})}\BibitemShut {NoStop}%
\bibitem [{\citenamefont {Melikhov}\ and\ \citenamefont
  {Simula}(2002)}]{Melikhov:2001pm}%
  \BibitemOpen
  \bibfield  {author} {\bibinfo {author} {\bibfnamefont {D.}~\bibnamefont
  {Melikhov}}\ and\ \bibinfo {author} {\bibfnamefont {S.}~\bibnamefont
  {Simula}},\ }\href {\doibase 10.1103/PhysRevD.65.094043} {\bibfield
  {journal} {\bibinfo  {journal} {Phys. Rev.}\ }\textbf {\bibinfo {volume}
  {D65}},\ \bibinfo {pages} {094043} (\bibinfo {year} {2002})}\BibitemShut
  {NoStop}%
\bibitem [{\citenamefont {Choi}\ and\ \citenamefont {Ji}(2004)}]{Choi:2004ww}%
  \BibitemOpen
  \bibfield  {author} {\bibinfo {author} {\bibfnamefont {H.-M.}\ \bibnamefont
  {Choi}}\ and\ \bibinfo {author} {\bibfnamefont {C.-R.}\ \bibnamefont {Ji}},\
  }\href {\doibase 10.1103/PhysRevD.70.053015} {\bibfield  {journal} {\bibinfo
  {journal} {Phys. Rev.}\ }\textbf {\bibinfo {volume} {D70}},\ \bibinfo {pages}
  {053015} (\bibinfo {year} {2004})}\BibitemShut {NoStop}%
\bibitem [{\citenamefont {Jaus}(2003)}]{Jaus:2002sv}%
  \BibitemOpen
  \bibfield  {author} {\bibinfo {author} {\bibfnamefont {W.}~\bibnamefont
  {Jaus}},\ }\href {\doibase 10.1103/PhysRevD.67.094010} {\bibfield  {journal}
  {\bibinfo  {journal} {Phys. Rev.}\ }\textbf {\bibinfo {volume} {D67}},\
  \bibinfo {pages} {094010} (\bibinfo {year} {2003})}\BibitemShut {NoStop}%
\bibitem [{\citenamefont {Krassnigg}(2001)}]{KrassniggDiss:2001}%
  \BibitemOpen
  \bibfield  {author} {\bibinfo {author} {\bibfnamefont {A.}~\bibnamefont
  {Krassnigg}},\ }\href@noop {} {Ph.D. thesis},\ \bibinfo  {school}
  {Karl-Franzens University of Graz} (\bibinfo {year} {2001})\BibitemShut
  {NoStop}%
\bibitem [{\citenamefont {Bagdasaryan}\ \emph {et~al.}(1985)\citenamefont
  {Bagdasaryan}, \citenamefont {Esaibegian},\ and\ \citenamefont
  {Ter-Isaakian}}]{Bagdasaryan:1984kz}%
  \BibitemOpen
  \bibfield  {author} {\bibinfo {author} {\bibfnamefont {A.~S.}\ \bibnamefont
  {Bagdasaryan}}, \bibinfo {author} {\bibfnamefont {S.~V.}\ \bibnamefont
  {Esaibegian}}, \ and\ \bibinfo {author} {\bibfnamefont {N.~L.}\ \bibnamefont
  {Ter-Isaakian}},\ }\href@noop {} {\bibfield  {journal} {\bibinfo  {journal}
  {Yad. Fiz.}\ }\textbf {\bibinfo {volume} {42}},\ \bibinfo {pages} {440}
  (\bibinfo {year} {1985})}\BibitemShut {NoStop}%
\bibitem [{\citenamefont {Samsonov}(2003)}]{Samsonov:2003hs}%
  \BibitemOpen
  \bibfield  {author} {\bibinfo {author} {\bibfnamefont {A.}~\bibnamefont
  {Samsonov}},\ }\href@noop {} {\bibfield  {journal} {\bibinfo  {journal} {J.
  High Energy Phys.}\ }\textbf {\bibinfo {volume} {12}},\ \bibinfo {pages}
  {061} (\bibinfo {year} {2003})}\BibitemShut {NoStop}%
\bibitem [{\citenamefont {Aliev}\ and\ \citenamefont
  {Savci}(2004)}]{Aliev:2004uj}%
  \BibitemOpen
  \bibfield  {author} {\bibinfo {author} {\bibfnamefont {T.~M.}\ \bibnamefont
  {Aliev}}\ and\ \bibinfo {author} {\bibfnamefont {M.}~\bibnamefont {Savci}},\
  }\href {\doibase 10.1103/PhysRevD.70.094007} {\bibfield  {journal} {\bibinfo
  {journal} {Phys. Rev.}\ }\textbf {\bibinfo {volume} {D70}},\ \bibinfo {pages}
  {094007} (\bibinfo {year} {2004})}\BibitemShut {NoStop}%
\bibitem [{\citenamefont {Bhagwat}\ and\ \citenamefont
  {Maris}(2008)}]{Bhagwat:2006pu}%
  \BibitemOpen
  \bibfield  {author} {\bibinfo {author} {\bibfnamefont {M.~S.}\ \bibnamefont
  {Bhagwat}}\ and\ \bibinfo {author} {\bibfnamefont {P.}~\bibnamefont
  {Maris}},\ }\href {\doibase 10.1103/PhysRevC.77.025203} {\bibfield  {journal}
  {\bibinfo  {journal} {Phys. Rev.}\ }\textbf {\bibinfo {volume} {C77}},\
  \bibinfo {pages} {025203} (\bibinfo {year} {2008})}\BibitemShut {NoStop}%
\bibitem [{\citenamefont {Hawes}\ and\ \citenamefont
  {Pichowsky}(1999)}]{Hawes:1998bz}%
  \BibitemOpen
  \bibfield  {author} {\bibinfo {author} {\bibfnamefont {F.~T.}\ \bibnamefont
  {Hawes}}\ and\ \bibinfo {author} {\bibfnamefont {M.~A.}\ \bibnamefont
  {Pichowsky}},\ }\href {\doibase 10.1103/PhysRevC.59.1743} {\bibfield
  {journal} {\bibinfo  {journal} {Phys. Rev.}\ }\textbf {\bibinfo {volume}
  {C59}},\ \bibinfo {pages} {1743} (\bibinfo {year} {1999})}\BibitemShut
  {NoStop}%
\bibitem [{\citenamefont {de~Melo}\ and\ \citenamefont
  {Frederico}(1997)}]{deMelo:1997hh}%
  \BibitemOpen
  \bibfield  {author} {\bibinfo {author} {\bibfnamefont {J.~P. B.~C.}\
  \bibnamefont {de~Melo}}\ and\ \bibinfo {author} {\bibfnamefont
  {T.}~\bibnamefont {Frederico}},\ }\href {\doibase 10.1103/PhysRevC.55.2043}
  {\bibfield  {journal} {\bibinfo  {journal} {Phys. Rev.}\ }\textbf {\bibinfo
  {volume} {C55}},\ \bibinfo {pages} {2043} (\bibinfo {year}
  {1997})}\BibitemShut {NoStop}%
\bibitem [{\citenamefont {Gurtler}\ \emph {et~al.}(2008)\citenamefont {Gurtler}
  \emph {et~al.}}]{Gurtler:2008zz}%
  \BibitemOpen
  \bibfield  {author} {\bibinfo {author} {\bibfnamefont {M.}~\bibnamefont
  {Gurtler}} \emph {et~al.} (\bibinfo {collaboration} {QCDSF Collaboration}),\
  }\href@noop {} {\bibfield  {journal} {\bibinfo  {journal} {PoS}\ }\textbf
  {\bibinfo {volume} {LATTICE2008}},\ \bibinfo {pages} {051} (\bibinfo {year}
  {2008})}\BibitemShut {NoStop}%
\bibitem [{\citenamefont {Garcia~Gudino}\ and\ \citenamefont
  {Toledo~Sanchez}()}]{Gudino:2013jaa}%
  \BibitemOpen
  \bibfield  {author} {\bibinfo {author} {\bibfnamefont {D.}~\bibnamefont
  {Garcia~Gudino}}\ and\ \bibinfo {author} {\bibfnamefont {G.}~\bibnamefont
  {Toledo~Sanchez}},\ }\href@noop {} {\ }\Eprint
  {http://arxiv.org/abs/1305.6345} {arXiv:1305.6345} \BibitemShut {NoStop}%
\bibitem [{\citenamefont {Braguta}\ and\ \citenamefont
  {Onishchenko}(2004)}]{Braguta:2004kx}%
  \BibitemOpen
  \bibfield  {author} {\bibinfo {author} {\bibfnamefont {V.~V.}\ \bibnamefont
  {Braguta}}\ and\ \bibinfo {author} {\bibfnamefont {A.~I.}\ \bibnamefont
  {Onishchenko}},\ }\href {\doibase 10.1103/PhysRevD.70.033001} {\bibfield
  {journal} {\bibinfo  {journal} {Phys. Rev.}\ }\textbf {\bibinfo {volume}
  {D70}},\ \bibinfo {pages} {033001} (\bibinfo {year} {2004})}\BibitemShut
  {NoStop}%
\bibitem [{\citenamefont {Biernat}\ \emph {et~al.}(2008)\citenamefont
  {Biernat}, \citenamefont {Fuchsberger}, \citenamefont {Schweiger},\ and\
  \citenamefont {Klink}}]{Biernat:2007dn}%
  \BibitemOpen
  \bibfield  {author} {\bibinfo {author} {\bibfnamefont {E.~P.}\ \bibnamefont
  {Biernat}}, \bibinfo {author} {\bibfnamefont {K.}~\bibnamefont
  {Fuchsberger}}, \bibinfo {author} {\bibfnamefont {W.}~\bibnamefont
  {Schweiger}}, \ and\ \bibinfo {author} {\bibfnamefont {W.~H.}\ \bibnamefont
  {Klink}},\ }\href {\doibase 10.1007/s00601-008-0316-5} {\bibfield  {journal}
  {\bibinfo  {journal} {Few Body Syst.}\ }\textbf {\bibinfo {volume} {44}},\
  \bibinfo {pages} {311} (\bibinfo {year} {2008})}\BibitemShut {NoStop}%
\bibitem [{\citenamefont {Bakker}\ and\ \citenamefont
  {Biernat}(2010)}]{Bakker:2010}%
  \BibitemOpen
  \bibfield  {author} {\bibinfo {author} {\bibfnamefont {B.~L.~G.}\
  \bibnamefont {Bakker}}\ and\ \bibinfo {author} {\bibfnamefont {E.~P.}\
  \bibnamefont {Biernat}},\ }\href@noop {} {\bibfield  {journal} {\bibinfo
  {journal} {PoS}\ }\textbf {\bibinfo {volume} {LC2010}},\ \bibinfo {pages}
  {015} (\bibinfo {year} {2010})}\BibitemShut {NoStop}%
\bibitem [{\citenamefont {Walecka}(1974)}]{ref.01}%
  \BibitemOpen
  \bibfield  {author} {\bibinfo {author} {\bibfnamefont {J.~D.}\ \bibnamefont
  {Walecka}},\ }\href {\doibase 10.1016/0003-4916(74)90208-5} {\bibfield
  {journal} {\bibinfo  {journal} {Annals Phys.}\ }\textbf {\bibinfo {volume}
  {83}},\ \bibinfo {pages} {491} (\bibinfo {year} {1974})}\BibitemShut
  {NoStop}%
\bibitem [{\citenamefont {Kupelwieser}\ and\ \citenamefont
  {Schweiger}()}]{Kupelwieser:2013nqa}%
  \BibitemOpen
  \bibfield  {author} {\bibinfo {author} {\bibfnamefont {D.}~\bibnamefont
  {Kupelwieser}}\ and\ \bibinfo {author} {\bibfnamefont {W.}~\bibnamefont
  {Schweiger}},\ }\href@noop {} {\ }\Eprint {http://arxiv.org/abs/1312.0863}
  {arXiv:1312.0863} \BibitemShut {NoStop}%
\bibitem [{\citenamefont {Gomez-Rocha}\ \emph {et~al.}(2013)\citenamefont
  {Gomez-Rocha}, \citenamefont {Schweiger},\ and\ \citenamefont
  {Senekowitsch}}]{Gomez-Rocha:2013zma}%
  \BibitemOpen
  \bibfield  {author} {\bibinfo {author} {\bibfnamefont {M.}~\bibnamefont
  {Gomez-Rocha}}, \bibinfo {author} {\bibfnamefont {W.}~\bibnamefont
  {Schweiger}}, \ and\ \bibinfo {author} {\bibfnamefont {O.}~\bibnamefont
  {Senekowitsch}},\ }\href@noop {} {\  (\bibinfo {year} {2013})},\ \Eprint
  {http://arxiv.org/abs/1311.1936} {arXiv:1311.1936 [hep-ph]} \BibitemShut
  {NoStop}%
\end{thebibliography}%

\end{document}